\documentclass[conference]{IEEEtran}
\IEEEoverridecommandlockouts
\usepackage{cite}
\usepackage{amsmath,amssymb,amsfonts}
\usepackage{algorithmic}
\usepackage{graphicx}
\usepackage{textcomp}
\usepackage{xcolor}
\def\BibTeX{{\rm B\kern-.05em{\sc i\kern-.025em b}\kern-.08em
    T\kern-.1667em\lower.7ex\hbox{E}\kern-.125emX}}
    
\usepackage{subfiles}
\usepackage{paralist}
\usepackage{amsmath,amsfonts,amsthm}
\usepackage{mathrsfs}
\usepackage{amsbsy}
\usepackage[mathscr]{eucal} %
\usepackage{algorithmic}
\DeclareMathOperator*{\argmax}{argmax}
\usepackage[noend,ruled,vlined,linesnumbered,resetcount]{algorithm2e}
\usepackage{booktabs}
\usepackage{graphicx}
\usepackage[T1]{fontenc}
\usepackage{textcomp}
\usepackage{xcolor}
\usepackage{comment}
\usepackage{hyperref}

\usepackage{multirow}
\usepackage{tabularx,stackengine,collcell}
\let\endminwd\relax
\newcolumntype{L}[1]{>{\collectcell\xminwd l{#1}}l<{\endminwd\endcollectcell}}
\newcolumntype{C}[1]{>{\collectcell\xminwd c{#1}}c<{\endminwd\endcollectcell}}
\newcolumntype{R}[1]{>{\collectcell\xminwd r{#1}}r<{\endminwd\endcollectcell}}
\def\minwd#1#2#3\endminwd{\stackengine{0pt}{#3}{\rule{#2}{0pt}}{O}{#1}{F}{F}{L}}
\newcommand\xminwd[1]{\minwd#1}

\usepackage[center]{subfigure}
\usepackage{bbold}
\usepackage{mathtools}
\usepackage{enumitem}
\usepackage{bbm}

\newtheorem{theorem}{Theorem}
\newtheorem{lemma}{Lemma}

\SetAlFnt{\small}
\SetKwComment{Comment}{$\triangleright$\ }{}

\SetCommentSty{mycommfont}
\SetKwInput{KwData}{Input}
\SetKwInput{KwResult}{Output}
\newcommand{\method}{\textsc{Slugger}\xspace}

\newcommand{\sweg}{\textsc{SWeG}\xspace}

\newcommand{\mosso}{\textsc{MoSSo}\xspace}

\newcommand{\sags}{\textsc{SAGS}\xspace}

\newcommand{\randomized}{\textsc{Randomized}\xspace}

\newcommand{\apx}{\textsc{Apxmdl}\xspace}

\newcommand{\kGs}{\textsc{K-Gs}\xspace}

\newcommand{\ssl}{\textsc{S2L}\xspace}

\newcommand{\ssumm}{\textsc{SSumM}\xspace}

\newcommand{\lone}{$\ell_{1}$\xspace}

\newcommand{\original}{$G = (V, E)$\xspace}

\newcommand{\summary}{$\tilde{G} = (S, P, C^+, C^-)$\xspace}

\newcommand{\sG}{$\overline{G} = (S, P^+, P^-, H)$\xspace}

\newcommand{\sN}{$S$\xspace}

\newcommand{\hieredge}{$H$\xspace}

\newcommand{\pP}{$P^+$\xspace}

\newcommand{\pM}{$P^-$\xspace}

\newcommand{\rootS}{$R$\xspace}

\newcommand\blue[1]{\textcolor{blue}{#1}}

\newtheoremstyle{problemstyle}  %
{3pt}                                               %
{3pt}                                               %
{\normalfont}                               %
{}                                                  %
{\bfseries\itshape}                 %
{\normalfont\bfseries:}         %
{.5em}                                          %
{}                                                  %
\theoremstyle{problemstyle}

\newtheorem{problem}{Problem}

\newcommand{\smallsection}[1]{{\vspace{0.05in} \noindent {\bf{\underline{\smash{#1}}}}}}

\newcommand\red[1]{\textcolor{red}{#1}}

\newcommand\grey[1]{\textcolor{gray}{#1}}

\def\endthebibliography{%
	\def\@noitemerr{\@latex@warning{Empty `thebibliography' environment}}%
	\endlist
}

\begin{document}

\title{SLUGGER: Lossless Hierarchical Summarization of Massive Graphs}

\author{
\IEEEauthorblockN{Kyuhan Lee\IEEEauthorrefmark{1}, Jihoon Ko\IEEEauthorrefmark{1}, and Kijung Shin}
\IEEEauthorblockA{\textit{Kim Jaechul Graduate School of AI, KAIST, Seoul, South Korea} \\
\{kyuhan.lee, jihoonko, kijungs\}@kaist.ac.kr} 
}

\maketitle
\begingroup\renewcommand\thefootnote{\IEEEauthorrefmark{1}}
\footnotetext{Equal contribution.}
\endgroup

\begin{abstract}
\textit{Given a massive graph, how can we exploit its hierarchical structure for concisely but exactly summarizing the graph? By exploiting the structure, can we achieve better compression rates than state-of-the-art graph summarization methods?}
		 
		The explosive proliferation of the Web has accelerated the emergence of large graphs, such as online social networks and hyperlink networks. Consequently, graph compression has become increasingly important to process such large graphs without expensive I/O over the network or to disk.
		Among a number of approaches, 
		\textit{graph summarization}, which in essence combines similar nodes into a \textit{supernode} and describe their connectivity concisely, protrudes with several advantages.
		However, we note that it fails to exploit pervasive hierarchical structures of real-world graphs as its underlying representation model enforces supernodes to be disjoint.		
		
		In this work, we propose the \textit{hierarchical graph summarization model}, which is an expressive graph representation model that includes the previous one proposed by Navlakha et al. 
		as a special case.
		The new model represents an unweighted graph using positive and negative edges between \textit{hierarchical supernodes}, each of which can contain others. 
		Then, we propose \method, a scalable heuristic for concisely and exactly representing a given graph under our new model.
		\method greedily merges nodes into supernodes while maintaining and exploiting their hierarchy, which is later pruned.
		\method significantly accelerates this process by sampling, approximation, and memoization.
		Our experiments on 16 real-world graphs show that \method is \textbf{(a) Effective:} yielding up to \textit{29.6\% more concise} summary
		than state-of-the-art lossless summarization methods, \textbf{(b) Fast:} summarizing a graph with \textit{$0.8$ billion edges} in a few hours,
		and \textbf{(c) Scalable:} scaling linearly with the number of edges in the input graph.
\end{abstract}

\section{Introduction}
\label{sec:intro}

Graphs are a natural and powerful abstraction for representing relations. %
Examples include social connections (i.e., social networks), citations between papers (i.e., citation networks), hyperlinks between webpages (i.e., hyperlink networks), and interactions between proteins (i.e., PPI networks). 

Due to the proliferation of the Web and its applications, which produce a large amount of data ceaselessly, massive graphs have emerged. Online social networks with over $20$ billion social connections \cite{dhulipala2016compressing,shin2019sweg}, hyperlink networks with $129$ billion hyperlinks \cite{JWS-0003}, and an online curation graph with over $100$ billion edges \cite{eksombatchai2018pixie} are well-suited examples. 

\begin{figure}[t]
    \centering
	\subfigure[Relative size of outputs (PR dataset)]{
	\label{fig:jewel:concise} 
		\includegraphics[width= 0.55\linewidth]{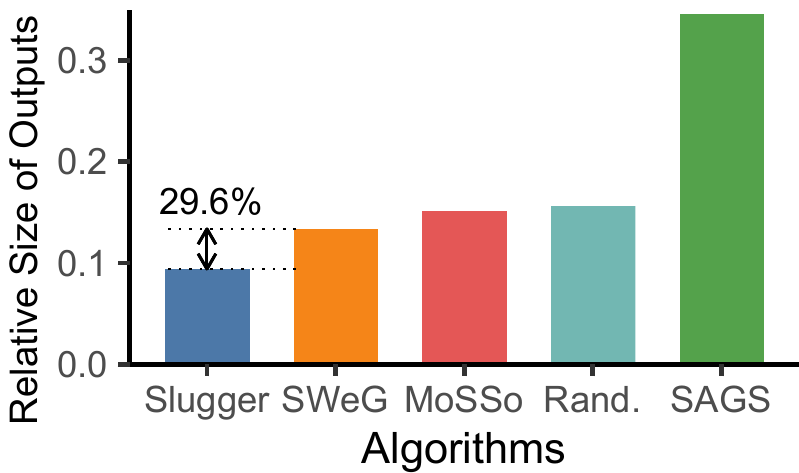}
	} 
	\subfigure[Scalability]{
	    \label{fig:scalability}
		\includegraphics[width= 0.35\linewidth]{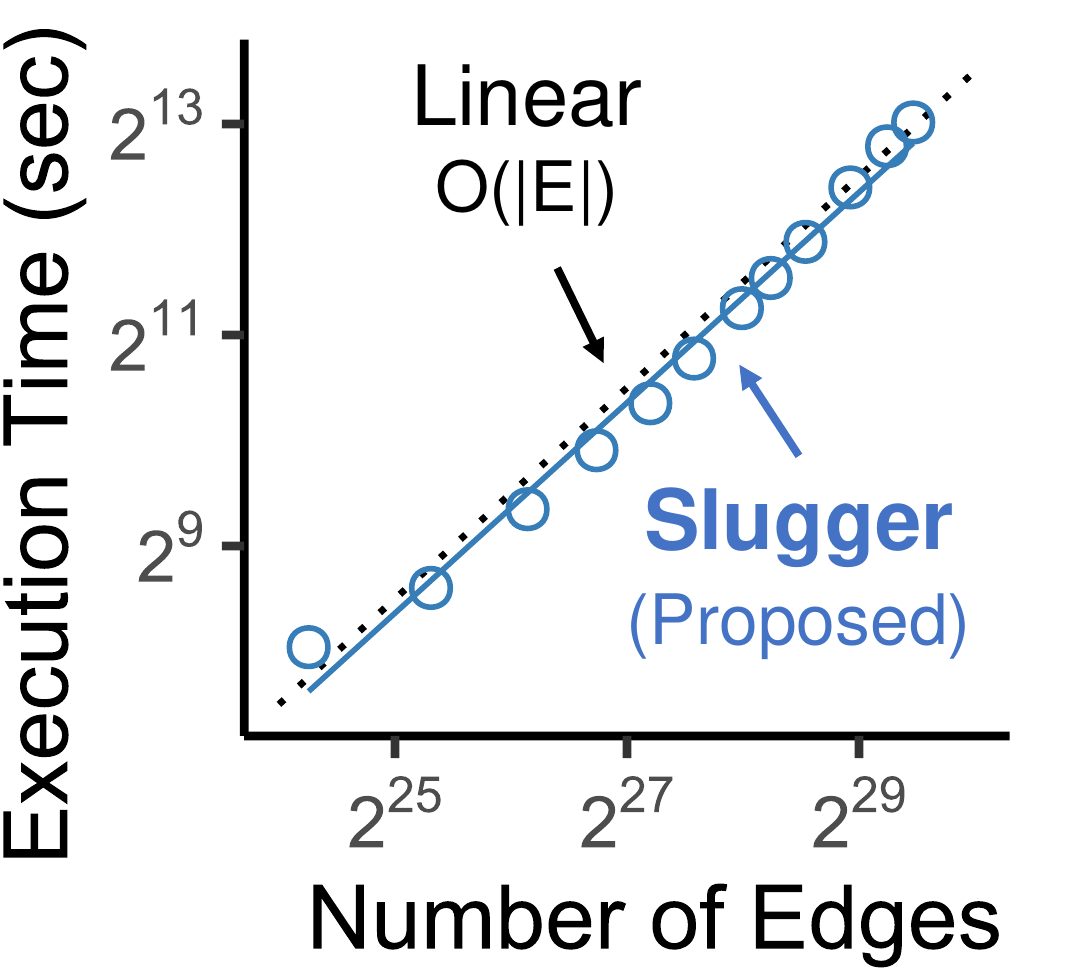}
	} 
	\caption{\label{fig:jewel} 
		\underline{\smash{Strengths of \method.}} (a) \method achieves up to $29.6\%$ better graph compression than its best competitor with similar speed. (b) \method scales linearly with the size of the input graph. See Sect. \ref{sec:experiments} for details.
	}
\end{figure}

Efficiently storing such large-scale graphs has become a critical issue since typical graph algorithms silently assume that the input graph fits in main memory, without causing I/O delays to the disk or over the network. 
While several programming models \cite{kang2009pegasus,low2012distributed,malewicz2010pregel} are available for out-of-core and distributed graph processing,
only a small fraction of graph algorithms naturally fit the programming models.

Consequently, a number of graph compression techniques have been proposed.
Lossless techniques include the WebGraph framework \cite{boldi2004webgraph} with node relabeling schemes
\cite{chierichetti2009compressing, apostolico2009graph, dhulipala2016compressing, boldi2011layered}, graph summarization \cite{navlakha2008graph, beg2018scalable, shin2019sweg, ko2020incremental}, %
and pattern mining \cite{buehrer2008scalable, koutra2014vog}; and lossy techniques include 
lossy graph summarization \cite{lefevre2010grass, riondato2017graph, lee2020ssumm}, sampling \cite{leskovec2006sampling}, and sketching \cite{khan2017toward,zhao2011gsketch}.

The focus of this work is \textit{graph summarization} \cite{navlakha2008graph, beg2018scalable, shin2019sweg, ko2020incremental}, where nodes with similar connectivity are combined into a \textit{supernode} so that their connectivity can be encoded together to save bits. The unweighted input graph $G$ is represented by (a) positive edges between supernodes (i.e., sets of nodes), (b) positive edges between ordinary nodes, and (c) negative edges between ordinary nodes, which $G$ is exactly restored from.

Graph summarization has multiple merits.
First, its outputs are three graphs, and thus they can be further compressed using any graph-compression techniques.
In other words, lossless graph summarization can be used as a preprocess to give additional compression \cite{shin2019sweg}. Moreover, the neighbors of each node can be retrieved by decompressing on-the-fly a small fraction of an output representation. This enables a wide range of graph algorithms (e.g., DFS, PageRank, and Dijkstra's) to run directly on an output representations, without decompressing all of it, while this may take longer than running graph algorithms on uncompressed graphs (see Sect.~\ref{sec:appendix:query}).

Despite its effectiveness, we note that the underlying graph representation model of graph summarization has limited expressive power in that supernodes need to be disjoint.
In other words, each node should belong to exactly one supernode.\footnote{A supernode can be a singleton, which contains a single ordinary node.}
As a result, there is a limit to exploiting hierarchical structures for compression, while
hierarchical and more generally overlapping structures are known to be pervasive \cite{leskovec2010kronecker,girvan2002community,sales2007extracting,ahnert2013power}.
That is, in many real-world graphs, a group of nodes with similar connectivity have subgroups with higher similarity, which in turn have subgroups with even higher similarity.

\begin{table}[t]
	\small
	\begin{center}
		\caption{Frequently-used symbols and definitions.}
		\label{tab:symndef}
		\vspace{-4mm}
		\scalebox{0.83}{
		    \hspace{-2mm}
		    \renewcommand{\arraystretch}{1.2}
			\begin{tabular}{|l|l|}
				\hline 
				\textbf{Symbol}  & \textbf{Definition}\\
				\hline
				\hline 
				\original & input graph with subnodes $V$ and subedges $E$\\
				\sG & hierarchical graph summarization model \\
				\sN & set of supernodes \\
				\pP & set of  positive edges ($p$-edges) between supernodes\\
				\pM & set of negative edges ($n$-edges) between supernodes\\
				\hieredge & set of hierarchy edges ($h$-edges) between supernodes \\

				\hline
				$T$  &  given number of iterations\\ 
				$\theta(t)$  &  threshold at the $t$-th iteration\\
				$C_t$ &set of candidate sets at the $t$-th iteration\\   
				\rootS & set of root nodes \\
				$S_X$ & set of supernodes in the hierarchy tree rooted at $X$\\
				\hline
				$\binom{A}{2}$ & set of all possible size-$2$ subsets of a set $A$ \\
				\hline
			\end{tabular}
		}
	\end{center}
\end{table}

In this work, we propose the \textit{hierarchical graph summarization model}, which is an expressive graph representation model where a supernode may contain smaller supernodes, which in turn contain even smaller supernodes. 
This new model, which represents the unweighted input graph using positive and negative edges between supernodes, includes the aforementioned model as a special case. 
In brief, a positive edge between two supernodes indicates the edges between all pairs of ordinary nodes between them; and a negative edge between two supernodes indicates no edge between any pair of ordinary nodes between the two supernodes. More precisely, an edge between two ordinary nodes exists if and only if there are more positive edges than negative edges between the supernodes that they belong to.

Concisely representing a given graph using our new model, which we call \textit{lossless hierarchical graph summarization} also has the merits of graph summarization: natural combination with other graph-compression techniques and on-the-fly partial decompression (see Sect.~\ref{sec:appendix:query}).
Outputs are still graphs: (a) hierarchy trees of supernodes, (b) positive edges between supernodes, and (c) negative edges between supernodes.

Our algorithmic contribution is to propose \method (\textbf{S}calable \textbf{L}ossless S\textbf{u}mmarization of \textbf{G}raphs with Hi\textbf{er}archy), a scalable heuristic for concisely and exactly representing a given graph under our new model. That is, \method searches for smallest sets of (a) supernodes, (b) positive edges, and (c) negative edges that exactly represent the given graph.
To this end, it greedily merges nodes into supernodes while maintaining and exploiting their hierarchy, which is later pruned.
\method also significantly accelerates this process by sampling, approximation, and memoization.
As a result, it scales linearly with the number of edges in the input graph, and empirically, its output is consistently more concise compared to those of state-of-the-art graph summarization methods.

In summary, our contributions in this work are as follows:
\begin{itemize}[leftmargin=*]
	\item \textbf{New Graph Representation Model}: We propose the hierarchical graph summarization model, which generalizes the previous graph summarization model while inheriting its advantages.
	Our new model is capable of naturally and concisely representing hierarchical structures, which are pervasive in real-world graphs.
	\item \textbf{Fast and Effective Algorithm}: We propose \method for concisely and exactly representing a given graph under our new model. With linear scalability (Fig.~\ref{fig:scalability}), \method compresses a graph with $0.8$ billion edges within a few hours.   
	 \method achieves up to $29.6\%$ better compression than the best graph summarization methods (Fig.~\ref{fig:jewel:concise}).
	\item \textbf{Extensive Experiments}: We substantiate the superiority of \method over $4$ state-of-the-art graph summarization methods on $16$ real-world graphs from various domains.
\end{itemize}

\noindent\textbf{Reproducibility:} The source code and the datasets are available at \url{https://github.com/KyuhanLee/slugger}.

In Sect.~\ref{sec:preliminaries}, we present our new graph representation model and formally define the lossless hierarchical graph summarization problem.
In Sect.~\ref{sec:method}, we describe our proposed algorithm \method, and we analyze its time and space complexity. 
In Sect.~\ref{sec:experiments}, we share experimental results.
After reviewing related works in Sect.~\ref{sec:related}, we draw conclusions in Sect.~\ref{sec:conclusion}.

\begin{figure*}[t]
	\centering
	\includegraphics[width= \linewidth]{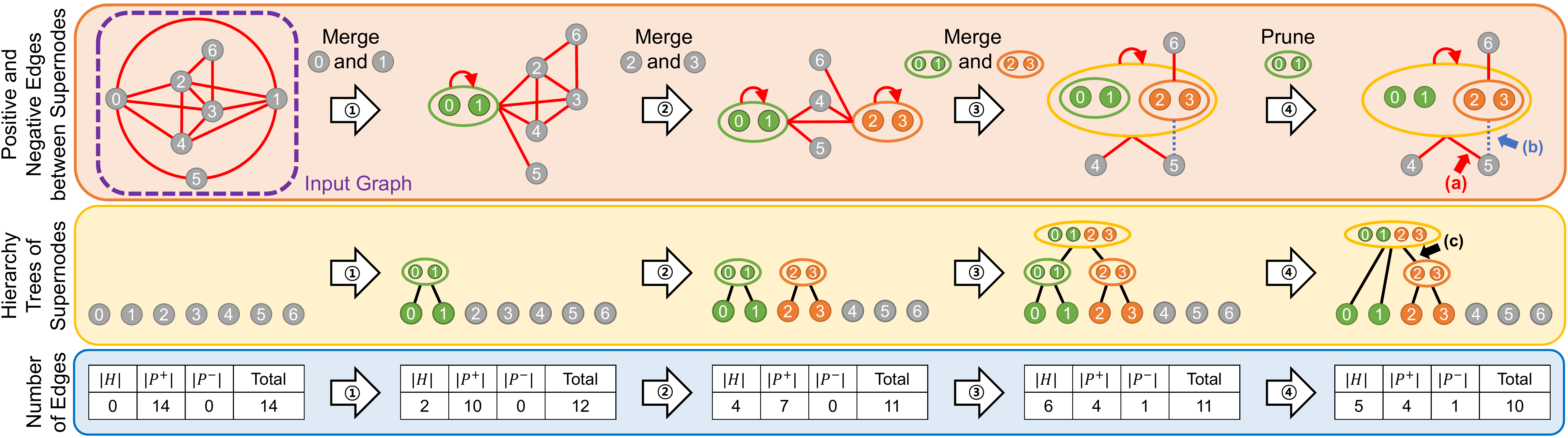}
	\caption{\label{fig:model} 
		\underline{\smash{Example procedure of hierarchical graph summarization.}} The figure shows how positive edges (\red{red}), negative edges (\blue{dotted blue}) (see the upper panel), hierarchy edges (see the middle panel), and the total count of edges (see the lower panel) change during the five steps of an example procedure.
		The procedure is lossless, and in every step, the three types of edges exactly describe the input graph
		(See Sect.~\ref{sec:model:hier} for how to interpret them).
		For example, in the last step, the positive edge (a) indicates that four edges $(0,5)$, $(1,5)$, $(2,5)$, $(3,5)$ exist in the input graph, and the negative edge (b) indicates that, among them, $(2,5)$ and $(3,5)$ should be removed. The hierarchy edge (c) indicates that the supernode $\{0,1,2,3\}$ contains the supernode $\{2,3\}$. Note that the total count of edges decreases from $14$ to $10$ as the procedure progresses.
	}
\end{figure*}
	
\section{Graph Representation Models}
\label{sec:preliminaries}

In this section, we first review the previous graph summarization model. 
Then, we present our new model, namely the hierarchical graph summarization model.
Lastly, we formally define the lossless hierarchical graph summarization problem.
We list some frequently-used symbols in Table~\ref{tab:symndef}.  

\smallsection{Input graph:}
We consider a simple undirected graph \original, where $V$ denotes the set of nodes and $E$ denotes the set of edges, and we use $(u, v)$ or $(v,u)$ to denote the undirected edge between two nodes $u, v \in V$. %
While both previous and proposed models and their algorithms can be easily extended to graphs with edge directions and/or self-loops, we focus on simple undirected graphs for simple descriptions of them. 
We call nodes and edges in $G$ subnodes and subedges, respectively, to distinguish them from supernodes, described below. 

\subsection{Graph Summarization Model}
\label{sec:model:prev}

In this subsection, we review the graph summarization model \cite{navlakha2008graph}, which is the underlying model of lossless graph summarization \cite{navlakha2008graph,khan2015set,shin2019sweg,ko2020incremental}.
In a nutshell, this model describes a given graph by (a) connections between groups of nodes and (b) exceptions.

\smallsection{Model description:}
The \textit{graph summarization model} \cite{navlakha2008graph}, which we denote by \summary, describes a given graph \original using (a) a set $P$ of edges between \textit{supernodes} $S$, where $S$ is a partition of $V$, (b) a set $C^+$ of \textit{positive edges} between subnodes, and (c) a set $C^-$ of \textit{negative edges} between subnodes.
Specifically, each edge between two supernodes (i.e., each edge in $P$) indicates the edges between all pairs of nodes in the two supernodes; and $C^+$ are the edges that are in $E$ but not described by $P$. Similarly, $C^-$ are the edges that are described by $P$ but not in $E$. 

\smallsection{Lossless graph summarization:}
The \textit{lossless graph summarization} problem is to find the parameters (i.e., $S$, $P$, $C^+$ and $C^-$) of the graph summarization model so that it represents a given graph concisely, minimizing $|P|+|C^+|+|C^-|$. %
Once $S$ is determined, finding the best $|P|$, $|C^+|$, and $|C^-|$ is trivial \cite{navlakha2008graph}. Thus the crux of the problem is to find the best $S$.
Intuitively, for conciseness, supernodes should be formed with subnodes with similar connectivity, and a number of heuristics \cite{navlakha2008graph,khan2015set,shin2019sweg,ko2020incremental} based on this idea have been developed.

\smallsection{Limitations:}
Hierarchical structures are pervasive in real-world graphs \cite{leskovec2010kronecker,girvan2002community,sales2007extracting}.
Essentially, in many real-world graphs, a group of subnodes with similar connectivity (e.g., students of a university) have subgroups with higher similarity (e.g., students of each department), which in turn have subgroups with even higher similarity (e.g., students advised by the same advisor).
However, the graph summarization model described above is not designed to effectively exploit such hierarchical structures because supernodes in the model are enforced to be disjoint. Note that the set $S$ of supernodes in the model should be a partition of $V$. That is, $\bigcup_{A \in S} A = V$ and $A \cap B = \emptyset$ for all $A,B\in S$.

\subsection{Hierarchical Graph Summarization Model}
\label{sec:model:hier}

In this subsection, we propose the hierarchical graph summarization model for exploiting pervasive hierarchical structures of real-world graphs.
To this end, the model allows supernodes to be hierarchical. See Fig.~\ref{fig:model} for an example.

\smallsection{Model description - (1) components:}
The \textit{hierarchical graph summarization model}, which we denote by \sG, describes a given graph \original using (a) a set $H$ of hierarchy edges between supernodes $S$, (b) a set $P^+$ of \textit{positive edges} between supernodes, and (c) a set $P^-$ of \textit{negative edges} between supernodes. 
The sets $P^+$ and $P^-$ can contain only undirected edges and self-loops.
We denote the undirected edge between two supernodes $A\in S$ and $B\in S$ by $(A,B)$ or $(B,A)$, and we denote the self-loop at a supernode $A\in S$ by $(A,A)$.
The set $H$ contains directed edges between supernodes, whose implication is described below.
From now on, we call positive edges \textit{p-edges}, negative edges \textit{n-edges}, and edges in $H$ \textit{h-edges}, for simplicity.

Note that, unlike the previous model, each supernode (i.e., a subset of $V$) in $S$ may contain smaller supernodes, which in turn may contain even smaller supernodes.
Specifically, for all supernodes $A\neq B\in S$, either $A\cap B = \emptyset$, $A\subsetneq B$, or $B\subsetneq A$ should hold, and thus their hierarchical relations can be described by a forest of supernodes $S$, where a supernode $A$ is the parent of a supernode $B$ if and only if $A$ is the smallest proper superset of $B$.
The set $H$ corresponds to the set of edges, which are directed from a parent to a child, of the forest.
The supernodes $S$ are divided into three groups according to their position in the forest: (a) leaf nodes, (b) internal nodes, and (c) root nodes, according to their positions in $H$. Specifically, root nodes are nodes without parents, leaf nodes are those without children, and the others are internal nodes.
Note that leaf nodes are singleton supernodes consisting of a single subnode.

\smallsection{Model description - (2) interpretation:}
Intuitively, as shown in Fig.~\ref{fig:model}, a $p$-edge between two supernodes indicates the edges between all pairs of subnodes in the two supernodes; and an $n$-edge between two supernodes indicates no edge between any pair of subnodes in the two supernodes. More precisely, there exists an edge between two subnodes in the input graph $G$ if and only if there are more $p$-edges than $n$-edges between the supernodes that they belong to. Formally, there exists an edge between two subnodes $u$ and $v$ if and only if the following inequality holds:
\begin{multline*}
|\{(A,B)\in P^+ : u \in A , v \in B\}| \\ > |\{(A,B)\in P^- : u \in A , v \in B\}|.
\end{multline*}
As discussed in Sect.~\ref{sec:intro}, the neighbors of each node can be retrieved from our model  on-the-fly without converting all of it to the input graph. Pseudocode  and empirical results are provided in Sect.~\ref{sec:appendix:query}. This enables a wide range of graph algorithms (e.g., DFS, PageRank, and Dijkstra's) to run directly on our model. See Sect.~\ref{sec:appendix:algorithms} for examples.

\begin{figure*}[t]
	\centering
	\subfigure[$\Theta(nk)=o(n^{1.5})$ edges]{
		\label{inputgraph}
		\includegraphics[width= 0.23\linewidth]{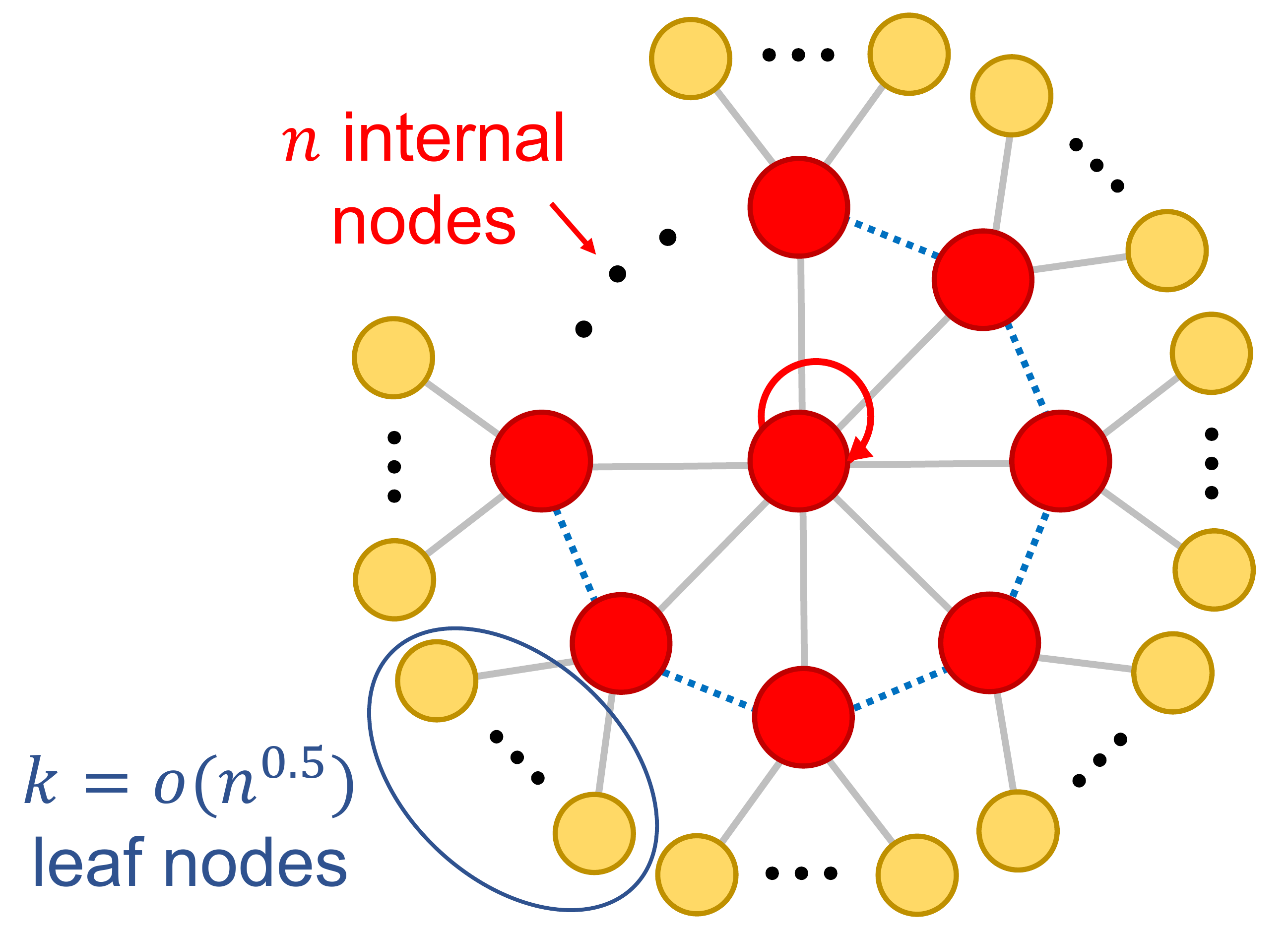}
	} 
	\subfigure[$\Theta(nk^2)=o(n^2)$ edges]{
		\label{adj}
		\includegraphics[width= 0.23\linewidth]{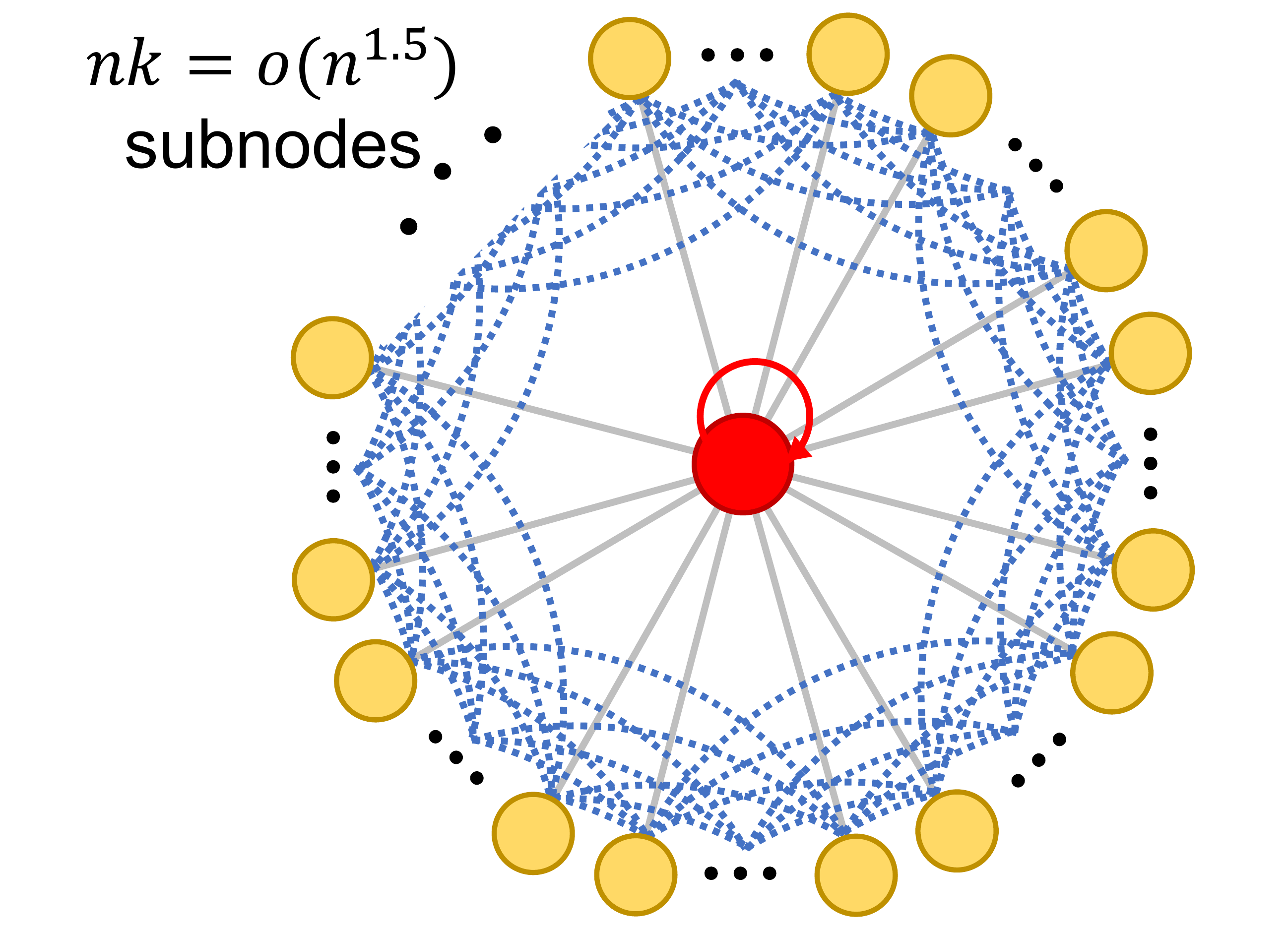}
	} 
	\subfigure[$\Theta(n^2 + nk)=\Theta(n^2)$ edges]{
		\label{summarygraph}
		\includegraphics[width= 0.23\linewidth]{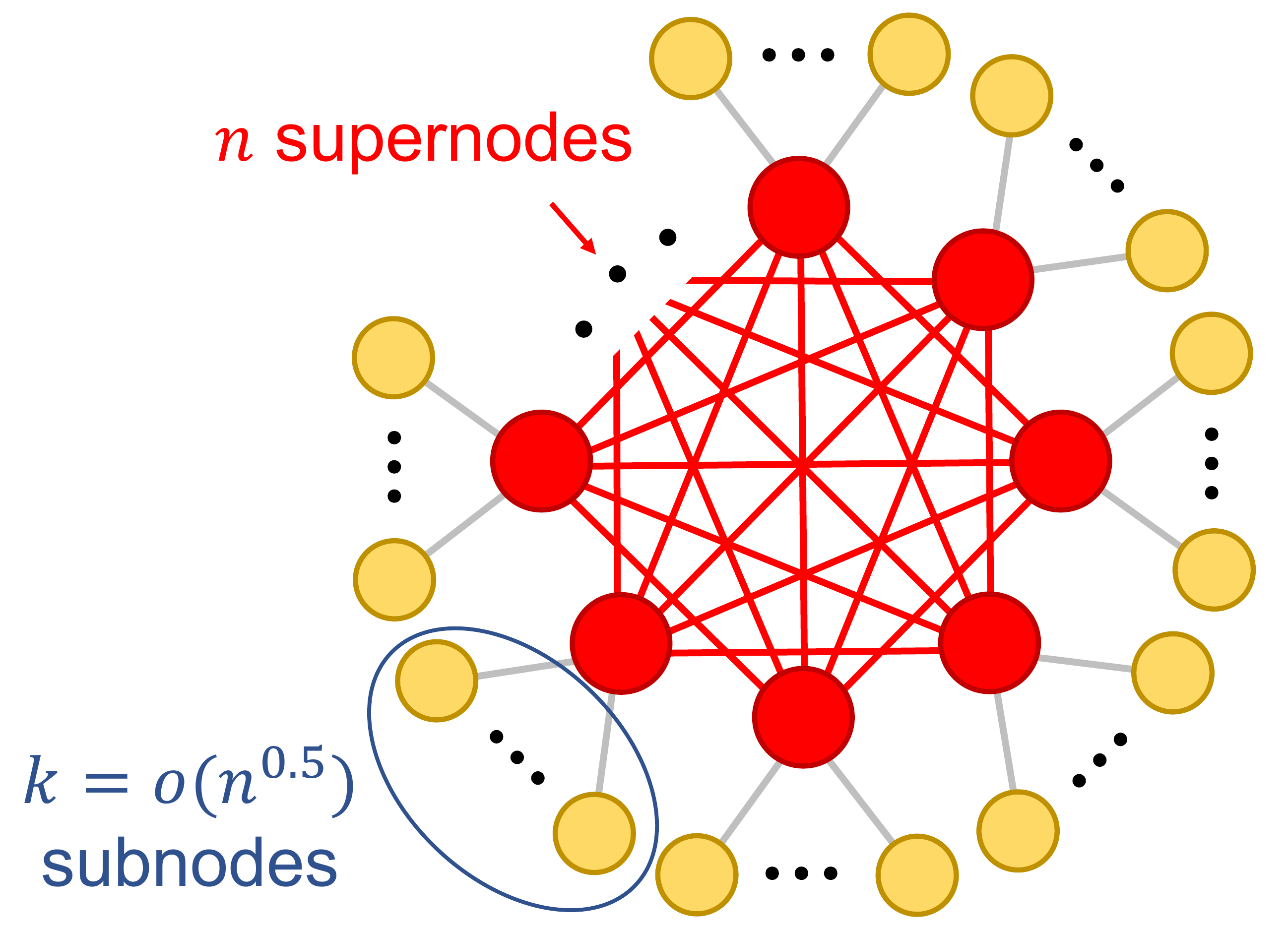}
	} 
	\subfigure[$\Theta(n^2k^2)=o(n^3)$ edges]{
		\label{readj}
		\includegraphics[width= 0.23\linewidth]{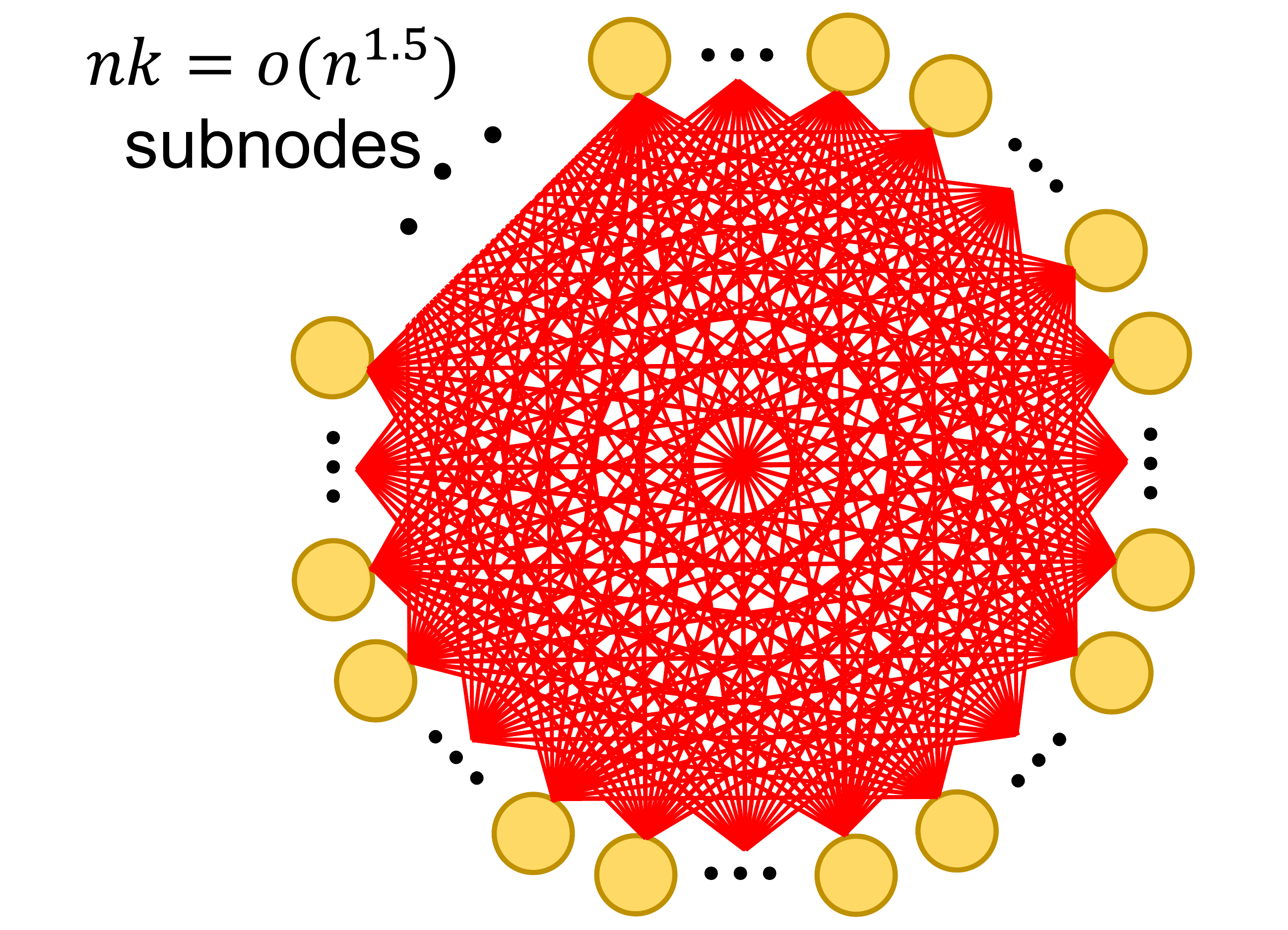}
	} \\ \vspace{-1mm}
	\caption{\label{fig:concept} 
		\underline{\smash{The hierarchical graph summarization model gives a concise representation.}}
		\red{Red} edges are p-edges, \blue{blue} edges are n-edges, and \grey{grey} edges are h-edges.
		Consider the graph $G$ represented by the hierarchical graph summarization model \sG in (a) with $o(n^{1.5})$ edges. 
		That is,  $|P^+|+|P^-|+|H|=o(n^{1.5})$.
		However, the graph cannot be represented  with $o(n^{1.5})$ edges by the previous summarization model \summary, as formalized in Theorem~\ref{theorem:nk}. (b), (c), and (d) show example instances of the previous summarization model with $o(n^2)$, $\Theta(n^2)$, and $o(n^3)$ edges, respectively.
	}
\end{figure*}

\smallsection{Comparison with the previous model:}
The hierarchical graph summarization model generalizes and thus includes as a special case the previous graph summarization model described in Sect.~\ref{sec:model:prev}. %
Specifically, $P$ in \summary can be expressed as positive edges between root nodes in our model; and $C^+$ and $C^-$ can be expressed as positive and negative edges between singleton supernodes in our model.
In other words, if we use the terms in our model, in the previous model, all positive edges are restricted to be only between root nodes or between singleton nodes, and all negative edges are restricted to be only between singleton nodes.
Thus, the hierarchical graph summarization model can always describe a given graph more concisely than or at least as concise as the previous model can.
Moreover, in Fig.~\ref{fig:concept} and Theorem~\ref{theorem:nk}, we provide an example where our model can be strictly more concise than the previous model.
\begin{theorem}[Conciseness of the Hierarchical Graph Summarization Model: an Example]\label{theorem:nk}	
    Consider the graph $G$ represented by the hierarchical graph summarization model \sG  in Fig.~\ref{fig:concept}(a) with $o(n^{1.5})$ edges. That is,  $|P^+|+|P^-|+|H|\in o(n^{1.5})$.
    The graph $G$ cannot be represented with $o(n^{1.5})$ edges by the previous graph summarization model, described in Sect.~\ref{sec:model:prev}. That is, $|C^+|+|C^-|+|P| \notin o(n^{1.5})$ for every \summary of $G$.
\end{theorem}

\begin{proof}
See Sect.~\ref{sec:appendix:example} for a proof.
\end{proof}

\subsection{Problem Definition}

The \textit{lossless hierarchical graph summarization} problem, which we consider in this work, is to find the parameters (i.e., $S$, $P^+$, $P^-$ and $H$) of the hierarchical graph summarization model so that it represents a given graph concisely.
Since the number of bits required for \sG is roughly proportional to the number of edges in it, we aim to minimize Eq.~\eqref{eq:encodingCost}, which we call the \textit{encoding cost}.
\begin{align}
Cost(\overline{G}) := |P^+| + |P^-| + |H|.
\label{eq:encodingCost}
\end{align}
The considered problem is formalized in Problem~\ref{problem}.
\begin{problem}[\label{problem}Lossless Hierarchical Graph Summarization]~
	\begin{itemize}[leftmargin=*]
		\item \textbf{Given:} an unweighted input graph \original
		\item \textbf{Find:} a hierarchical summary graph \sG
		\item \textbf{to Minimize:} $Cost(\overline{G})=|P^+| + |P^-| + |H|$.
	\end{itemize}
\end{problem}
The hardness of Problem~\ref{problem} and also the hardness of the lossless graph summarization problem \cite{navlakha2008graph} are open.

\section{Proposed Algorithm: \method}
\label{sec:method}

In this section, we propose \method (\textbf{S}calable \textbf{L}ossless S\textbf{u}mmarization of \textbf{G}raphs with Hi\textbf{er}archy), a scalable heuristic for Problem~\ref{problem}.
That is, given a graph $G$, it \textit{encodes} $G$ with the hierarchical graph summarization model while minimizing the \textit{encoding cost} (i.e., Eq.~\eqref{eq:encodingCost}).
It performs a randomized greedy search based on three main ideas:
\begin{compactitem}[$\bullet$]
    \item \method greedily merges supernodes and updates $h$-edges, $p$-edges and $n$-edges simultaneously, reducing the encoding cost in every merger.
    \item \method accelerates finding the best encoding in each merger by memoizing the best ones in the previous mergers. Moreover, \method rapidly and effectively samples promising node pairs to be merged. As a result, it achieves linear scalability. 
    \item \method further reduces the encoding cost without any information loss by pruning supernodes that do not contribute to succinct encoding.
\end{compactitem}

In Sect.~\ref{sec:costfunction}, we present the cost functions used in \method. In Sect.~\ref{sec:overview}, we give an overview of \method and then describe each step in detail. In Sect.~\ref{sec:analysis}, we analyze its time and space complexity.
\subsection{Cost Functions of \method}
\label{sec:costfunction}
In this subsection, we introduce the cost functions used in each step of greedy search by \method.
To this end, we first divide the encoding cost (i.e., Eq.~(\ref{eq:encodingCost})) into two as follows:
\begin{equation}
Cost(\overline{G}) = Cost^H(\overline{G}) + Cost^P(\overline{G}), \label{eq:costDivide}
\end{equation}
where $Cost^H(\overline{G}):=|H|$ is the encoding cost for the hierarchy between supernodes, and $Cost^P(\overline{G}):=|P^+|+|P^-|$ is that for $p$-edges and $n$-edges. %

We let $R$ be the set of root nodes (see Sect.~\ref{sec:model:hier}) in $\overline{G}$; and let $\Pi_{R} :={R \choose 2} \cup\{(A,A):A\in R\}$ be all possible unordered pairs of root nodes\footnote{$\binom{R}{2}$ is the set of all possible size-$2$ subsets of $R$.}. 
We also let $S_X$ be the set consisting of a root node $X$ and its descendants in its hierarchy tree.
Then, we divide $Cost^H(\overline{G})$ into that for each root node as: %
\begin{equation}
Cost^{H}(\overline{G}) = \sum\nolimits_{A \in R} Cost^H_A(\overline{G}), \label{eq:hDivide} 
\end{equation}
where $Cost^H_A(\overline{G})$ denotes the number of h-edges in the hierarchy tree rooted at $A$.
We also divide $Cost^P(\overline{G})$ into that for each root node pair as: %
\begin{equation}
Cost^{P}(\overline{G}) = \sum\nolimits_{(A,B) \in \Pi_R} Cost^P_{A, B}(\overline{G}), \label{eq:pDivide1} 
\end{equation}
where $Cost^{P}_{A, B}(\overline{G})$ denotes the number of $p$-edges and $n$-edges between $S_A$ and $S_B$.
Based on the concept, in Eq.~\eqref{eq:pDivide2}, we define $Cost^{P}_A(\overline{G})$ as the number of $p$-edges and $n$-edges incident to any supernode in $S_A$.
\begin{equation}
Cost^{P}_A(\overline{G}) := \sum\nolimits_{X \in R} Cost^P_{A, X}(\overline{G}). \label{eq:pDivide2} 
\end{equation}
Then, we define the \textit{encoding cost for each root node} $A$ as: %
\begin{equation}
Cost_A(\overline{G}) := Cost^H_A(\overline{G}) + Cost^P_A(\overline{G}). \label{eq:costA} 
\end{equation}

The cost $Cost_A(\overline{G})$ for each root node $A$ and the cost $Cost^P_{A, B}(\overline{G})$ for each unordered pair $\{A,B\}$ are used in \method for decision making, as described below.

\subsection{Description of \method}
\label{sec:overview}
\begin{algorithm}[t]
	\SetAlgoLined
	\LinesNumbered
	\KwData{
		(a) input graph: \original  \\
		\hspace{9.7mm} (b) the number of iterations: $T$}
	\KwResult{hier. graph summ. model: \sG}
	$S\leftarrow \{\{u\}:u\in V \}$ \hspace{-2mm}\label{alg:line:init1}\Comment*[f]{initialize $\overline{G}$ to $G$} \\
	$R\leftarrow \{\{u\}:u\in V \}$  \\
	$P^+\leftarrow \{(\{u\}, \{v\}) : (u,v) \in E\}$, \\
	$P^-\leftarrow \emptyset$, \ \label{alg:line:init2} 
	$H \leftarrow \emptyset$, \ 
	$t$ $\leftarrow$ 1 \\ %
	\While{$ t \leq T$}{
		generate candidate sets $C_t\subseteq 2^{R}$ \label{alg:line:candidate} \Comment*[f]{Sect.~\ref{alg:candiate}} \\
		\For{$\mathbf{each}$ candidate set $D\in C_t$ \label{alg:line:merge1}}{
			merge some supernodes within $D$ and update $S$, $P^+$ $P^{-}$, $H$, and $R$ accordingly \label{alg:line:merge2} \Comment*[f]{Sect. \ref{alg:merging}}
		}
		$t$ $\leftarrow$ $t + 1$\\
	}
	prune  \sG \label{alg:line:pruning} \Comment*[f]{Sect.~\ref{alg:pruning}} \\
	{\bf return} \sG
	\caption{Overview of \method \label{algo:main}}
\end{algorithm}

Based on the cost functions we have defined, in this subsection, we describe \method, a scalable algorithm for Problem~\ref{problem}.
We first provide an overview and then describe each step in detail.

\subsubsection{\bf Overview (Algorithm~\ref{algo:main})}\label{sec:method:search:overview}
Given an input graph \original and the number of iterations $T$, \method aims to find a hierarchical graph \sG of $G$ that minimizes the encoding cost $Cost(\overline{G})$ (i.e., Eq.~\eqref{eq:encodingCost}).
\method first initializes $\overline{G}$ to $G$. That is, it sets $S = \{\{u\} : u \in V\}$, $P^+ = \{(\{u\}, \{v\}) : (u,v) \in E\}$, $P^-=\emptyset$, and $H=\emptyset$. Then, \method repeatedly merges supernodes and at the same time updates their encoding. To this end, it alternatively runs the following two steps $T$ times:

\begin{compactitem}[$\bullet$]
	\item \textbf{Candidate generation (line~\ref{alg:line:candidate}):} 
	 Naively finding a root node pair whose merger gives the largest reduction in the encoding cost needs to take $O{|R| \choose 2}$ pairs into consideration.
	 This step aims to speed up this process by dividing root nodes into smaller candidate sets. Each candidate set consists of root nodes whose merger is likely to bring a reduction in the encoding cost.

	\item \textbf{Merging (lines~\ref{alg:line:merge1}-\ref{alg:line:merge2}):} \method greedily merges a root node pair among those sampled within each candidate set, which is obtained in the previous step. Simultaneously, \method updates $p$-edges and $n$-edges incident to the merged nodes and/or their $1$-level descendants by exploiting the hierarchy between supernodes.
	\method accelerates this update through memoization.
\end{compactitem}
After repeating alternatively the above steps $T$ times, \method executes the following step:
\begin{compactitem}[$\bullet$]
	\item \textbf{Pruning (line~\ref{alg:line:pruning}):} \method further reduces the encoding cost by pruning supernodes that do not contribute to concise encoding.
	Note that there is no information loss, as $p$-edges, $n$-edges, and $h$-edges are updated accordingly.
	
\end{compactitem}
After being pruned, the current hierarchical graph summarization model \sG is returned as the output of \method. 
Below, we describe each step in detail.

\subsubsection{\bf Candidate Generation Step}\label{alg:candiate} 

In this step, \method divides root nodes into candidate sets within which \method searches for root node pairs to be merged with a large reduction in the encoding cost.
In order for the search to be rapid and effective,
each candidate set should be small and consisting of root nodes whose merger is likely to give a reduction in the encoding cost.
We note that merging two root nodes whose distance is three or larger always increases the encoding cost, as formalized in Lemma~\ref{lemma:3hop}.

\begin{lemma}[Unpromising Pairs]\label{lemma:3hop}	
	Let the distance between two root nodes $A\in R$ and $B\in R$ be the number of the edges in the shortest path in $G$ between any subnode in $A$ and that in $B$.
	If the distance between $A$ and $B$ is $3$ or larger, then
	\begin{equation}
	Cost_{A}(\overline{G})+Cost_{B}(\overline{G}) 
- Cost^{P}_{A, B}(\overline{G}) <  Cost_{A\cup B}(\hat{G}), \label{eq:3hop}
	\end{equation}
	where $\hat{G}$ is the hierarchical graph summarization model that is updated from $\overline{G}$ as $A$ and $B$ are merged, as described in Sect.~\ref{alg:merging}.
	
\end{lemma}
\begin{proof}
See Sect.~\ref{sec:appendix:3hop} for a proof.
\end{proof}

Thus, we group root nodes within distance $2$ as a candidate set, and to this end, we use min-hashing as in \sweg~\cite{shin2019sweg}.
Specifically, \method iteratively divides root nodes using shingle values at most $10$ times and then randomly so that each candidate set consists of at most $500$ nodes (see \cite{supple} for the effect of the size).  %
By using different random seeds at each iteration, \method varies the candidate sets so that more root node pairs can be considered to be merged.

\begin{algorithm}[t]
	\SetAlgoLined
	\LinesNumbered
	\KwData{(a) input graph: \original \\
		\quad (b) curr. hier. graph summ. model: \sG  \\
		\quad (c) current iteration number: $t$ \\
		\quad (d) set of root nodes: $R$ \\
		\quad (e) candidate root node set: $D$ \\
	}
	\KwResult{updated \sG}
	\caption{Merging Step\label{alg:merge}}
	$Q$ $\leftarrow$ $D$ \label{alg:line:skip1}  \\
	\While{$|Q| > 1$}{
		pick and remove a random root node $A$ from $Q$ \\
		$B$ $\leftarrow$ $\argmax_{Z \in Q} Saving(A, Z, \overline{G})$ \\ %
		\If{$Saving(A,B) \geq \theta (t)$ \label{alg:line:threshold}}{
			$\overline{G}$ $\leftarrow$ $Merge\_and\_Update(A, B, \overline{G})$\label{alg:line:merge} %
			\Comment*[f]{Sect.~\ref{alg:merging}} \\
			$R$ $\leftarrow$ $(R \setminus \{A, B\}) \cup \{A \cup B\}$ \\
			$Q$ $\leftarrow$ $(Q \setminus \{B\}) \cup \{A \cup B\}$
		}
	}
	{\bf return} \sG
\end{algorithm}

\subsubsection{\bf Merging Step}\label{alg:merging} 
In this step, \method repeats greedily merging root nodes and at the same time updates the encoding accordingly.
We first give an outline of this step and then describe the update process in detail.

\smallsection{Outline (Algorithm~\ref{alg:merge}):}
For each candidate set $D$ obtained above, \method randomly repeats selecting one root node $A$ and finds $B$ that maximizes Eq.~\eqref{eq:relative_saving} within $D$.
\begin{align}
Saving&  (A,B,\overline{G}) := \nonumber \\  
	& 1- \frac{Cost_{A \cup B}(\hat{G})}{Cost_{A}(\overline{G})+Cost_{B}(\overline{G})-Cost^{P}_{A, B}(\overline{G})},
\label{eq:relative_saving}
\end{align}
where $\hat{G}$ is the hierarchical graph summarization model that is updated from $\overline{G}$ as $A$ and $B$ are merged, as described later in this subsection.
Recall that $Cost_A(\overline{G})$ is the encoding cost for $A$ (see Eq.~(\ref{eq:costA}) for details). 
Thus, the denominator in Eq.~(\ref{eq:relative_saving}) is the encoding cost for $A$ and $B$ before their merger, and the numerator is the encoding cost for the merged supernode $A\cup B$ after their merger.

Then, as in \cite{shin2019sweg}, \method compares Eq.~\eqref{eq:relative_saving} with the merging threshold $\theta(t)$ in Eq.~\eqref{eq:threshold} at the current iteration $t$.
\begin{align}
\theta(t):=\left\{
\begin{array}{ll}
(1 + t)^{-1}  & \text{if} ~ t < T\\ 
0 & \text{if}~ t = T,
\end{array}
\right.
\label{eq:threshold}
\end{align}
Note that, at the beginning, this merging threshold is high, and thus it prevents merging selected root node pairs which do not reduce the encoding cost much, so that pairs with a larger reduction can be merged first. However, as the iteration proceeds, the threshold decreases, and the generated candidate set $D$ is exploited more.

After that, if the pair $\{A, B\}$ exceeds the merging threshold, \method merges them and updates the encoding accordingly, as described later in this subsection.
The process so far is repeated for every root node $A$ in $D$, and then the entire process is repeated for every candidate set $D$ obtained in the candidate generation step.

\smallsection{\bf Update of encoding:}
As presented above, \method repeatedly merges two root nodes, and additionally, in order to compute Eq.~\eqref{eq:relative_saving} for a candidate pair, the pair needs to be temporarily merged.
When two root nodes are merged, in addition to $h$-edges, $p$-edges and $n$-edges need to be updated if the encoding cost can be reduced by exploiting the new root node.
However, even when the hierarchy between supernodes is assumed to be fixed, it is still computationally expensive to exactly minimize the encoding cost.
Thus, when two root nodes are (temporarily) merged, \method relies on a sub-optimal encoding obtained by focusing on a small number of supernodes.

Consider two root nodes $A$ and $B$ are (temporarily) merged.
For each root node $X$, 
we let $\overline{S}_X$ be a set consisting of $X$ and its direct children.
\method updates \textbf{(Case 1)} $p$-edges and $n$-edges within $\{A\cup B\}\cup \overline{S}_{A} \cup \overline{S}_{B}$ and \textbf{(Case 2)} those between $\{A\cup B\}\cup \overline{S}_{A} \cup \overline{S}_{B}$ and $\overline{S}_C$ for each root node $C$ with such a $p$-edge or $n$-edge.

\begin{figure}[t]
    \vspace{-1.5mm}
    \includegraphics[width=\linewidth]{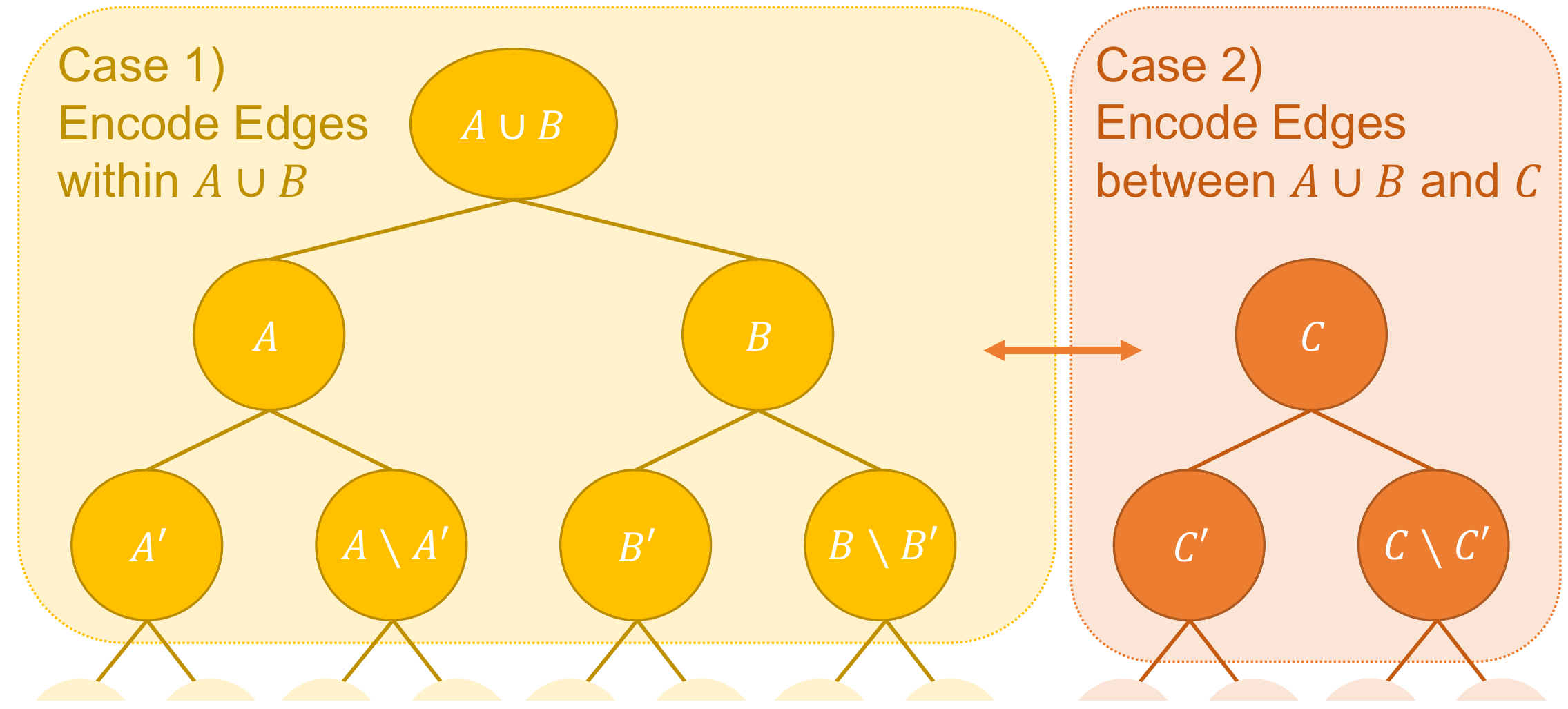}
    \caption{\label{fig:memoization} 
		\underline{\smash{Update of encoding in \method.}} When two root nodes $A$ and $B$ are merged, \method locally but rapidly updates encoding.
		\method updates $p$-edges and $n$-edges within the yellow panel, and then it updates $p$-edges and $n$-edges between the yellow and orange panels for each root node $C$ with such an edge.
		}
\end{figure}

\begin{compactitem}[$\bullet$]
\item \textbf{(Case 1)} \method updates $p$-edges and $n$-edges within $\{A\cup B\}\cup \overline{S}_{A} \cup \overline{S}_{B}$ (i.e., within the yellow panel in Fig.~\ref{fig:memoization} where the maximum number of supernodes is $7$), while fixing the other $p$-edges and $n$-edges.
Since we consider the pairs between at most $7$ supernodes, there are a constant number of possibilities, and a valid (i.e., representing the input graph $G$) one reducing the encoding cost most among them can be exhaustively searched.
\item \textbf{(Case 2)} \method updates $p$-edges and $n$-edges between $\{A\cup B\}\cup \overline{S}_{A} \cup \overline{S}_{B}$ and $\overline{S}_C$ (i.e., between the yellow and orange panels in Fig.~\ref{fig:memoization} where the maximum number of supernodes is $7$ and $3$, respectively), while fixing the other $p$-edges and $n$-edges, for each root node $C$ with such a $p$-edge or $n$-edge. 
Again, there are a constant number of possibilities, as we consider pairs between a set of size $7$ or smaller and a set of size $3$ or smaller. Thus, a valid (i.e., representing the input graph $G$) one reducing the encoding cost most among them can be exhaustively searched.
\end{compactitem}
When there are multiple best encodings, \method does not choose one immediately but chooses one later considering the right next step when $A\cup B$ (Cases 1 and 2) or $C$ (Case 2) is further merged.

Additionally, in both cases, \method further reduces the number of possibilities by focusing on those where 
$|\{(A, B)\in P^+ : u \in A , v \in B\}| - |\{(A, B)\in P^- : u \in A , v \in B\}|$ is either $1$ or $0$ for every two subnodes $u$ and $v$ in $G$ (see Sect.~\ref{sec:model:hier} for how this formula is used for interpreting the hierarchical graph summarization model).

\smallsection{\bf Memoization:}
While the number of possible output encodings in each of the two cases above 
(i.e., possible $p$-edges and $n$-edges between up to $10$ supernodes in Fig.~\ref{fig:memoization} after the update) is a constant,
it is inefficient to search all possibilities repeatedly.
\method memoizes the best encodings for all possible input cases (i.e., $p$-edges and $n$-edges between up to $10$ supernodes in Fig.~\ref{fig:memoization} before the update), whose number is also a constant,
when first exhaustively searching the possibilities in each of the two cases above.
Thus, it does not have to repeat the exhaustive search.
Note that the sets of possible input and output encodings and the best output encoding for each input encoding do not depend on the rest of the input graph if we limit our attention to up to $10$ supernodes in Fig.~\ref{fig:memoization} as in \method (see \cite{supple} for example pairs of input encodings and best output encodings).
Thus, the memoized results are independent of the input graph, and they can even be used when summarizing different input graphs.
The time and space required for exhaustively searching the possibilities once and memoizing the results are constant, and empirically, they are negligible compared to those required for summarizing a graph with millions of edges or more. 
Specifically, in our setting (see Sect.~\ref{sec:experiments:settings}), memoizing the best encodings for all possible possibilities takes less than $2$ seconds, and the memoized results, which are stored in a look-up table (i.e., pre-calculated array), take up only about $56$KB. While it is possible to run \method without memoization, it becomes several orders of magnitude slower without memoization.

\subsubsection{\bf Pruning Step}\label{alg:pruning} In this step, \method further reduces the encoding cost by removing supernodes that do not contribute to succinct encoding. Note that this step does not change what the current hierarchical graph summarization model \sG represents, and thus it still represents the input graph $G$ even after this step. 
\method prunes a supernode if all incident $p$-edges, $n$-edges, and $h$-edges can be replaced while decreasing the total encoding cost $Cost(\overline{G})$, which is defined as $|P^+|+|P^-|+|H|$ (see Eq.~(\ref{eq:encodingCost})).
Note that, since supernodes are connected by h-edges, removing one immediately reduces the encoding cost by reducing $|H|$. 
This pruning step consists of three substeps:
\begin{compactitem}[$\bullet$]
\item \textbf{(Step 1)} For each non-leaf node (i.e., each non-leaf node in the hierarchy forest) $A$ that is not incident to any $p$ or $n$-edge,
\method removes $A$ and connects h-edges from its parent to each of $A$'s direct children. For each removed non-leaf node, $|H|$ and thus the total encoding cost decrease by $1$.
The pseudocode of this substep is given in Algorithm~\ref{alg:prune} in Sect.~\ref{sec:appendix:prune}.
 
\item \textbf{(Step 2)} 
For each root node $A$ with only one incident non-loop $p$ or $n$-edge, \method removes $A$. Let $B$ be the other end point of the incident edge.
Then, for each  of $A$'s direct children $C$, \method removes the different type of edge between $B$ and $C$, if such an edge exists, or adds the same type of edge between $B$ and $C$, otherwise.
For each removed root node, $|H|$ decreases by the number of its children, while $|P^+|+|P^-|$ increases by at most the number of its children - 1. Therefore, the total encoding cost decrease by at least $1$.
The pseudocode of this substep is given in Algorithm~\ref{alg:prune} in Sect.~\ref{sec:appendix:prune}.

\item \textbf{(Step 3)} 
Since the previous graph summarization model is a special case of our model, as described in Sect.~\ref{sec:model:hier}, we can partially use its encoding if it reduces the encoding cost of our model.
For the previous model, once the root nodes are fixed, the best encoding can be found in $O(|E|)$ time (see \cite{shin2019sweg}). 
For each adjacent root node pair $A$ and $B$, we compare the encoding cost for edges between $S_{A}$ and $S_{B}$ in both models and choose the encoding of one with a smaller encoding cost. Since the previous model does not allow $p$- or $n$-edges incident to internal nodes, this substep may make more supernodes be pruned in the previous substeps. Thus, these three substeps can be repeated a few times.
\end{compactitem}

\noindent In Sect.~\ref{sec:expr:pruning}, we empirically demonstrate the effectiveness of each pruning substep.

\begin{figure*}[t!]
	\centering
	\subfigure[Relative size of outputs (absolute numbers of edges and standard deviations are available in \cite{supple})]{\label{fig:size}
		\includegraphics[width=0.9\linewidth]{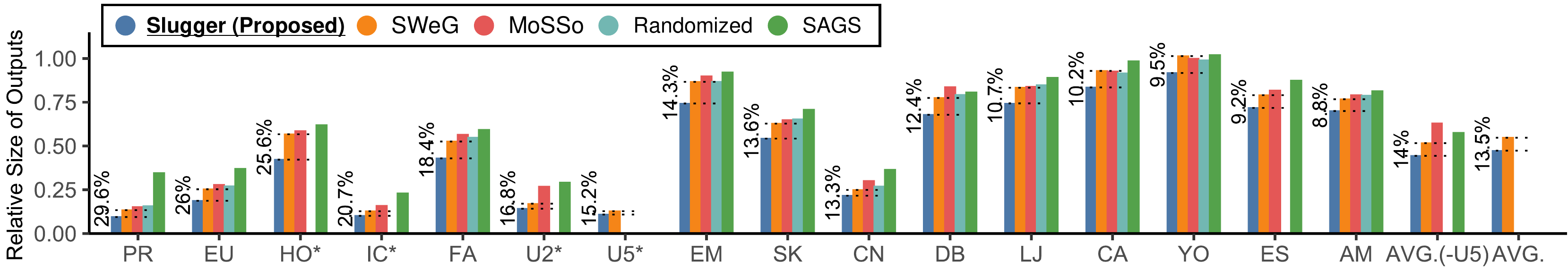}
	} \\
	\vspace{-2mm}
	\subfigure[Running time (the speed-ups of \method over \sweg and \sags are in orange and green, respectively)]{\label{fig:execution}
		\includegraphics[width=0.9\linewidth]{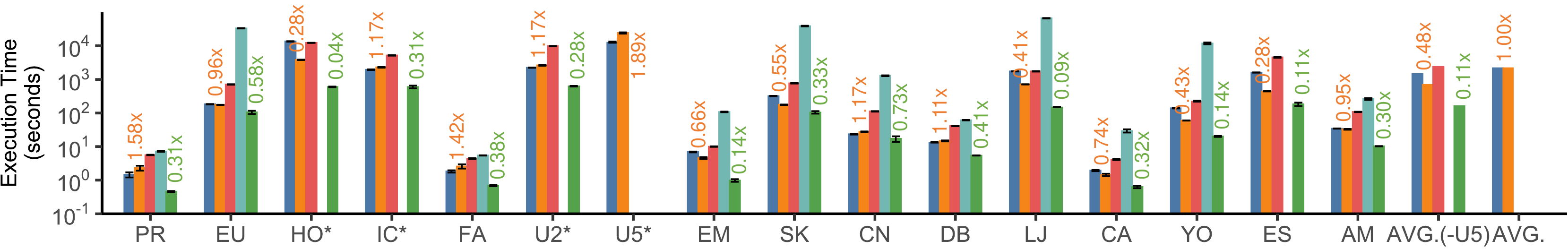}
	}
	\vspace{-1.9mm}
	\caption{\underline{\smash{\method provides concise representations of graphs.}} \label{fig:relativeOutput} 
	Missing competitors ran out of time ($>$24 hours) or out of memory ($>$128GB).
	We reported the means over five runs. For execution times, the error bars indicate $\pm1$ standard deviation. %
	Note that \textbf{\method yielded up to $\mathbf{29.6\%}$ smaller representations} than the best competitor \sweg with similar execution times.
	The datasets whose labels are marked with an asterisk are large ones with hundreds of million of edges.
	\method and \sweg successfully summarized the largest graph with about $\mathbf{0.8}$ billion edges, while all others failed.
	} %
\end{figure*}

\subsection{Complexity Analysis}\label{sec:analysis}

In this subsection, we analyze the time and space complexity of \method. 
We assume $|V| = O(|E|)$ for simplicity, and \textbf{all proofs can be found in Sect.~\ref{sec:appendix:proofs}}.

\smallsection{Time complexity:} The time complexity of memoization is $O(1)$ since it does not depend on the size of the input graph, and \method memoizes a constant number of cases. In our experimental settings (see Sect.~\ref{sec:experiments:settings}), the execution time for memoization did not exceed $2$ seconds. The other steps also take $O(|E|)$ time, as formalized in Lemmas~\ref{lemma:time:candigen}, \ref{lemma:time:merge} and \ref{lemma:time:pruning}.

\begin{lemma}[Time complexity of Candidate Generation Step \cite{shin2019sweg}] \label{lemma:time:candigen} The time complexity of the candidate generation step described in Sect.~\ref{alg:candiate} is $O(|E|)$.
\end{lemma}

\begin{lemma}[Time complexity of Merging Step] \label{lemma:time:merge} The time complexity of the merging step (i.e., Algorithm~\ref{alg:merge}) is $O(|E|)$. 
\end{lemma}

\begin{lemma}[Time Complexity of Pruning Step] \label{lemma:time:pruning} The overall time complexity of the pruning step described in Sect.~\ref{alg:pruning} is $O(|E|)$. 
\end{lemma}

Thus, the overall time complexity of \method (i.e., Algorithm~\ref{algo:main}), which repeats the candidate generation step and the merging step $T$ times, is $O(T \cdot |E|)$.
See Sect.~\ref{sec:exp:scalability} for empirical scalability.

\smallsection{Space complexity:} Due to the input graph, the space complexity of \method is at least $O(|E|)$. It is also an upper bound of the space complexity, as formalized in Lemma~\ref{lemma:space}.

\begin{lemma}[The Overall Space Complexity] The overall space complexity of Algorithm~\ref{algo:main} is $O(|E|)$. \label{lemma:space}
\end{lemma}

\section{Experiments}
\label{sec:experiments} 

We perform experiments to answer the following questions:
\begin{enumerate}[leftmargin=*]
	\item[Q1.] \textbf{Compactness}: Does \method yield a more compact summary than its state-of-the-art competitors? %
	\item[Q2.] \textbf{Speed \& Scalability}: Does \method scale linearly with the input graph's size? Can it summarize web-scale graphs?
	\item[Q3.] \textbf{Effects of Iterations}: How does the number of iterations $T$ affect the compression rates of \method?
	\item[Q4.] \textbf{Effectiveness of Pruning}: How does each pruning substep affect the compression rates of \method? %
	\item[Q5.] \textbf{Effects of the Height of Hierarchy Trees}: How do the heights of hierarchical trees affect the compression rates of \method? How large is the average depth of leaf nodes?
	\item[Q6.] \textbf{Composition of Outputs}: What is the proportion of edges of each type in the outputs of \method? 
\end{enumerate}
\subsection{Experimental Settings}
\label{sec:experiments:settings} 
\smallsection{Machines}: All experiments were conducted on a desktop with a $3.8$ GHz AMD Ryzen $3900$X CPU and $128$GB memory.

\begin{table}[t]
	\begin{center}
		\caption{\label{tab:DatasetTable} Real-world datasets used in our experiments.}
		\scalebox{0.9}{
		    \renewcommand{\arraystretch}{1.15}
			\begin{tabular}{|r|r|r|r|}
				\hline
				\textbf{Name} & \textbf{\# Nodes} & \textbf{\# Edges}  & \textbf{Summary}\\
				\hline
				\hline
				Caida (CA) & 26,475 & 53,381 & Internet\\
				Ego-Facebook (FA) & 4,039 & 88,234 & Social\\
				Protein (PR) & 6,229 & 146,160 & Protein Interaction\\
				Email-Enron (EM) & 36,692 & 183,831 & Email\\
				DBLP (DB) & 317,080 & 1,049,866 & Collaboration\\
				Amazon0601 (AM) & 403,394 & 2,443,408 & Co-purchase\\
				CNR-2000 (CN) & 325,557 & 2,738,969 & Hyperlinks\\
				Youtube (YO) & 1,134,890 & 2,987,624 & Social\\
				Skitter (SK) & 1,696,415 & 11,095,298 & Internet\\
				EU-05 (EU) & 862,664 & 16,138,468 & Hyperlinks\\
				Eswiki-13 (ES) & 970,327 & 21,184,931 & Social\\
				LiveJournal (LJ) & 3,997,962 & 34,681,189 & Social\\
				Hollywood (HO) & 1,985,306 & 114,492,816 & Collaboration\\
				IC-04 (IC) & 7,414,758 & 150,984,819 & Hyperlinks\\
				UK-02 (U2) & 18,483,186 & 261,787,258 & Hyperlinks\\
				UK-05 (U5) & 39,454,463 & 783,027,125 & Hyperlinks\\
				\hline
			\end{tabular}
		}
	\end{center}
\end{table}

\smallsection{Datasets}: We used $16$ real-world graphs listed in Table~\ref{tab:DatasetTable}. We removed all edge directions, duplicated edges, and self-loops.

\smallsection{Implementations}: We implemented \method and several state-of-the-art graph summarization algorithms in OpenJDK $12$: (a) \method where $T=20$ unless otherwise stated, (b) \randomized~\cite{navlakha2008graph}, (c) \sweg~\cite{shin2019sweg} where $T=20$ and $\epsilon = 0$ and (d) \sags~\cite{beg2018scalable} where $h=30$, $b= 10$, and $p= 0.3$. For (e) \mosso~\cite{ko2020incremental} where $e=0.3$ and $c= 120$, we used the Java implementation released by the authors. 

\smallsection{Evaluation Metrics}: Given a hierarchical graph summarization model \sG of a  graph \original, we measured
\begin{equation}
(|P^+| + |P^-| + |H|) ~/~ |E| \label{evaluation} 
\end{equation}
as the relative size of outputs. The numerator in Eq.~(\ref{evaluation}) is our objective $Cost(\overline{G})$ in Eq.~(\ref{eq:encodingCost}), and the denominator is the number of subedges in the input graph $G$. %
For a  previous model \summary,  Eq.~(\ref{evaluation}) is equivalent to 
\begin{equation}
(|P| + |C^+| + |C^-| + |H^*|)~/~|E|, \label{evaluation2} 
\end{equation}
where $H^*$ is the set of edges in hierarchy trees of height at most $1$ encoding which supernode each subnode belongs to.
Eq.~(\ref{evaluation}), Eq.~(\ref{evaluation2}), and runtimes were averaged over 5 trials, and the standard deviations were reported in \cite{supple}.

\subsection{Q1. Compactness}\label{sec:expr:compact}

We compared the size of the output representations obtained by \method and its competitors. As seen in Fig.~\ref{fig:relativeOutput}, \textbf{\method provided the most concise representations in all 16 considered datasets.} Especially, \method gave a $29.6\%$ smaller representation than the best competitor in the Protein dataset.

\subsection{Q2. Speed and Scalability}\label{sec:expr:speed} \label{sec:exp:scalability}

We evaluated the scalability of \method by measuring how its execution time grows as the number of edges in the input graph changes. We generated multiple graphs by sampling different numbers of nodes from the UK-05 dataset and used them. As seen in Fig.~\ref{fig:scalability}, \textbf{\method scaled linearly with the size of the input graph}.
This result is consistent with our analysis in Sect.~\ref{sec:analysis}.

As seen in Fig.~\ref{fig:execution}, in terms of speed, \method was comparable with \sweg, which consistently provided the second most concise representations. 
Specifically, \textbf{\method was faster than \sweg in 7 out of 16 datasets.} %
The theoretical time complexities of \method and \sweg are the same. However, if a similar number of pairs are merged, \method can be slower since it temporally encodes edges whenever nodes are merged, while \sweg encodes edges only once after the merging phase.
Moreover, \method has additional pruning steps.
While \sags was fastest in 15 out of 16 datasets, by rapidly deciding nodes to be merged, its output was least concise in the same number of datasets.
Only \method and \sweg successfully summarized the largest graph with about $\mathbf{0.8}$ billion edges, while all others failed.

\begin{table}[t]
	\centering
	\caption{\underline{\smash{The effects of the iteration number $T$ in \method.}} \label{fig:parameters} As $T$ increased, the outputs became concise. The compression rates (i.e., Eq.~\eqref{evaluation}) almost converged after $40$ iterations.} %
	\includegraphics[width=\linewidth]{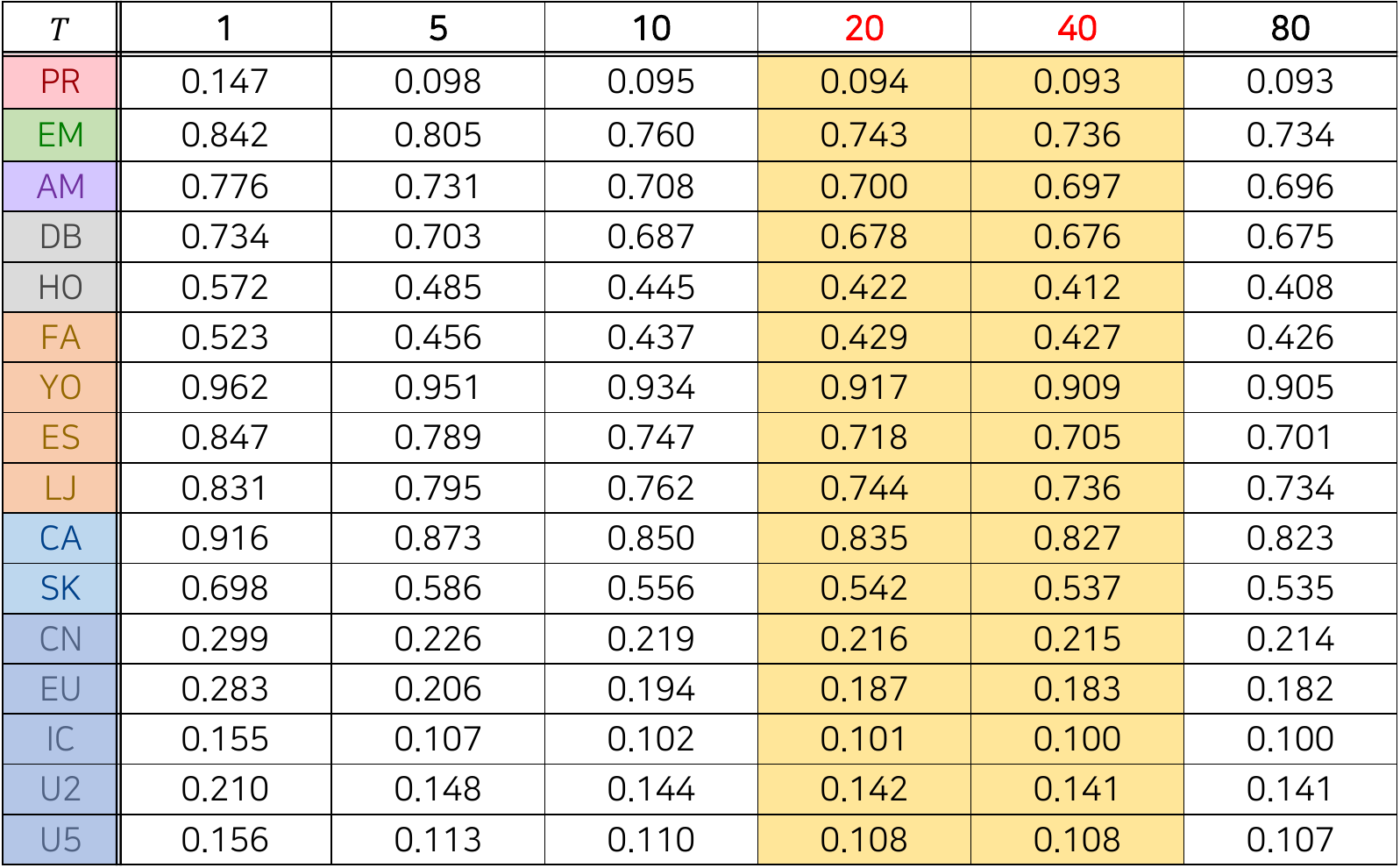} \\
\end{table}

\subsection{Q3. Effects of Iterations}\label{sec:expr:parameter}
We measured how the number of iteration $T$ affects the compactness of representations that \method gives. We changed $T$ from $1$ to $80$ on 16 datasets, and as seen in Table~\ref{fig:parameters}, the relative size of outputs decreased over iterations and almost converged after $40$ iterations. See \cite{supple} for the full results, including the effect of $T$ on speed.

\subsection{Q4. Effectiveness of Pruning}\label{sec:expr:pruning}

We measured how each substep of the pruning step of \method affects (a) the size of output representations, (b) the maximum height of hierarchy trees, and (c) the average depth of leaf nodes in \sG. 
We denote the state before using any pruning substep by $0$ and we denote the state after each $i$-th pruning substep by $i$ in Table~\ref{fig:pruning}.
Each substep reduced the size of output representations, the maximum height, and the average depth, justifying our design choices, while the first substep led to the largest reduction.
The full results are available in \cite{supple}.

\subsection{Q5. Effects of the Height of Hierarchy Trees}
\label{sec:appendix:height}

\begin{table}[t]
	\centering
	\caption{\underline{\smash{The pruning step in \method is effective.}} \label{fig:pruning} Every substep successfully decreased the size of output representations, proving the benefits of our design choices.}
	\includegraphics[width=\linewidth]{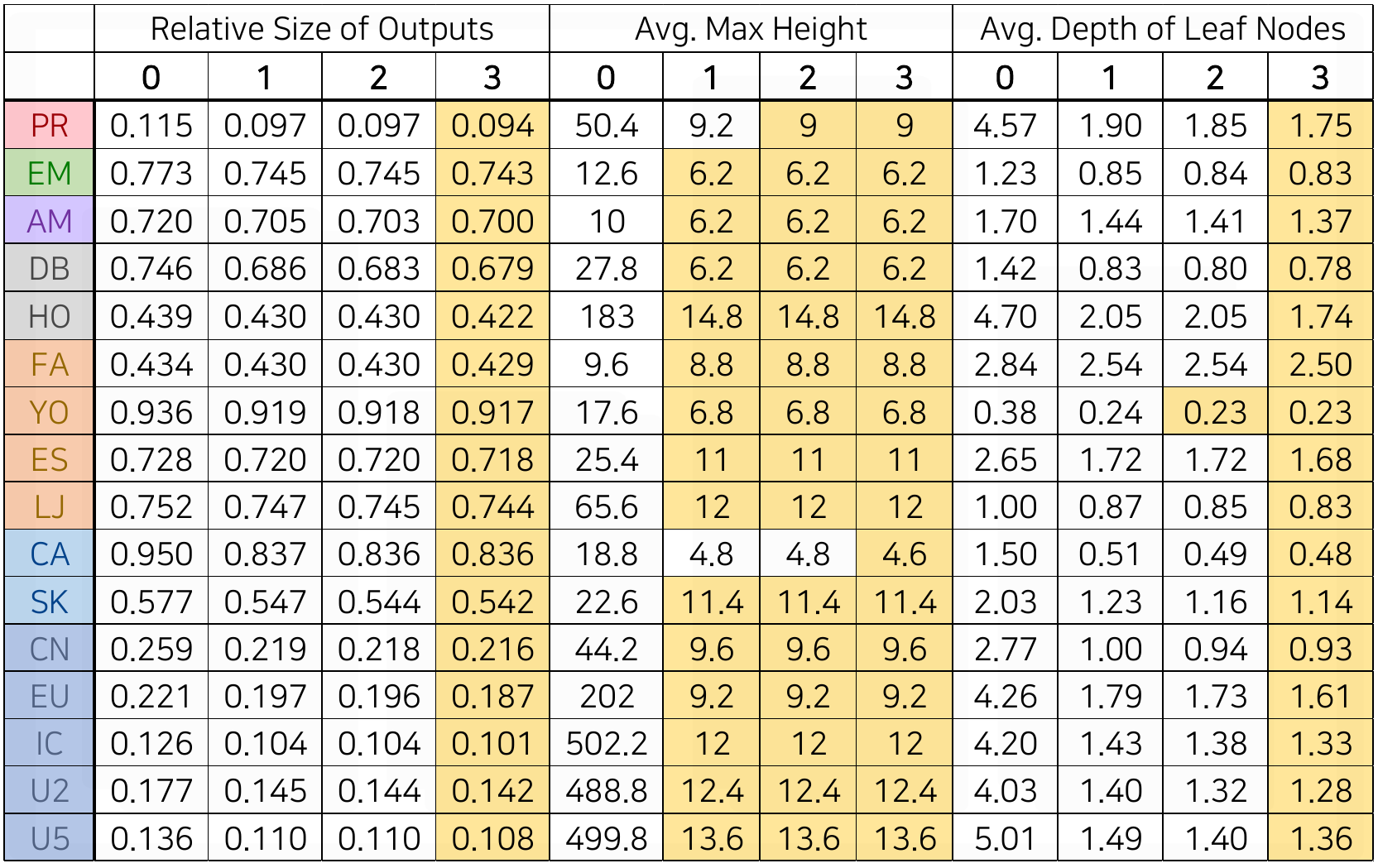} \\
\end{table}

In \method, the height of hierarchy trees can be as large as $(|V|-1)$, while the height is at most $1$ in all competitors. We investigated the choices between these two extremes. Specifically, we examined how the height of hierarchy trees affects the average depth of leaf nodes and the relative size of outputs.
To this end, we consider a variant of \method where the height never exceeds an upper bound $H_b$. In the variant, two supernodes are not merged, even when the merger decreases the size of outputs, if the merger results in a hierarchy tree with height greater than $H_b$.
As seen in Table~\ref{fig:height}, as the $H_b$ increased, the average depth of leaf nodes increased, while the relative size of outputs decreased.
The changes were gradual, and especially, the results at $H_b=10$ were close to the results in the original \method without any height limit. %
Notably, the average depth of leaf nodes was much lower than the upper bound $H_b$. The full results are available in \cite{supple}.

\subsection{Q6. Composition of Outputs}\label{sec:expr:edgeanalysis}

\begin{figure}[t]
    \centering
    \includegraphics[width=0.85\linewidth]{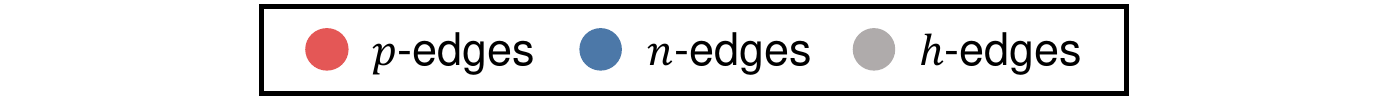}
    \includegraphics[width=\linewidth]{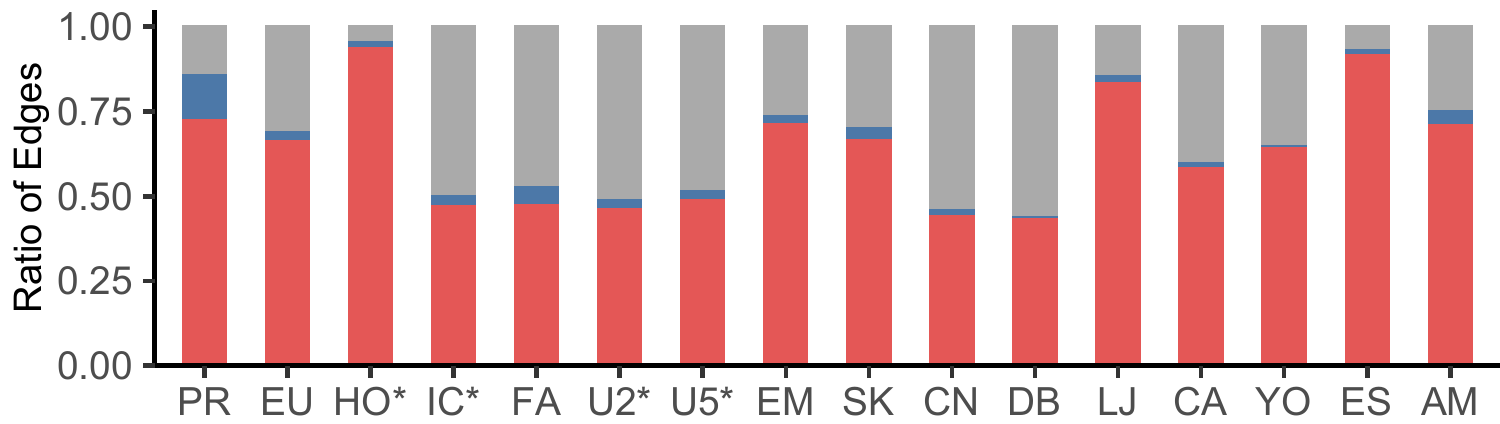}
    \caption{\underline{\smash{Edges of each type in outputs of \method.}} The proportion of $p$-edges was largest in $11$ datasets, and that of $h$-edges was largest in the remaining $5$ datasets.
    The proportion of $n$-edges was small in most datasets.
    }
    \label{fig:composition}
\end{figure}

We analyzed the proportion of edges of each type in the outputs of \method in Fig.~\ref{fig:composition}.
In $11$ datasets, $p$-edges accounted for the largest proportion,  and in the remaining $5$ datasets, $h$-edges accounted for the largest proportion.
The proportion of $n$-edges was less than $5.08\%$ in all the datasets except for the Protein dataset, where the proportion was $13.24\%$.

\begin{table}[t]
	\centering
	\caption{\underline{\smash{The effects of the height of hierarchy trees.}} As the upper bound $H_b$ of the height increased, the average depth of leaf nodes increased, and the relative size of outputs decreased.
	\label{fig:height}}
	\includegraphics[width=\linewidth]{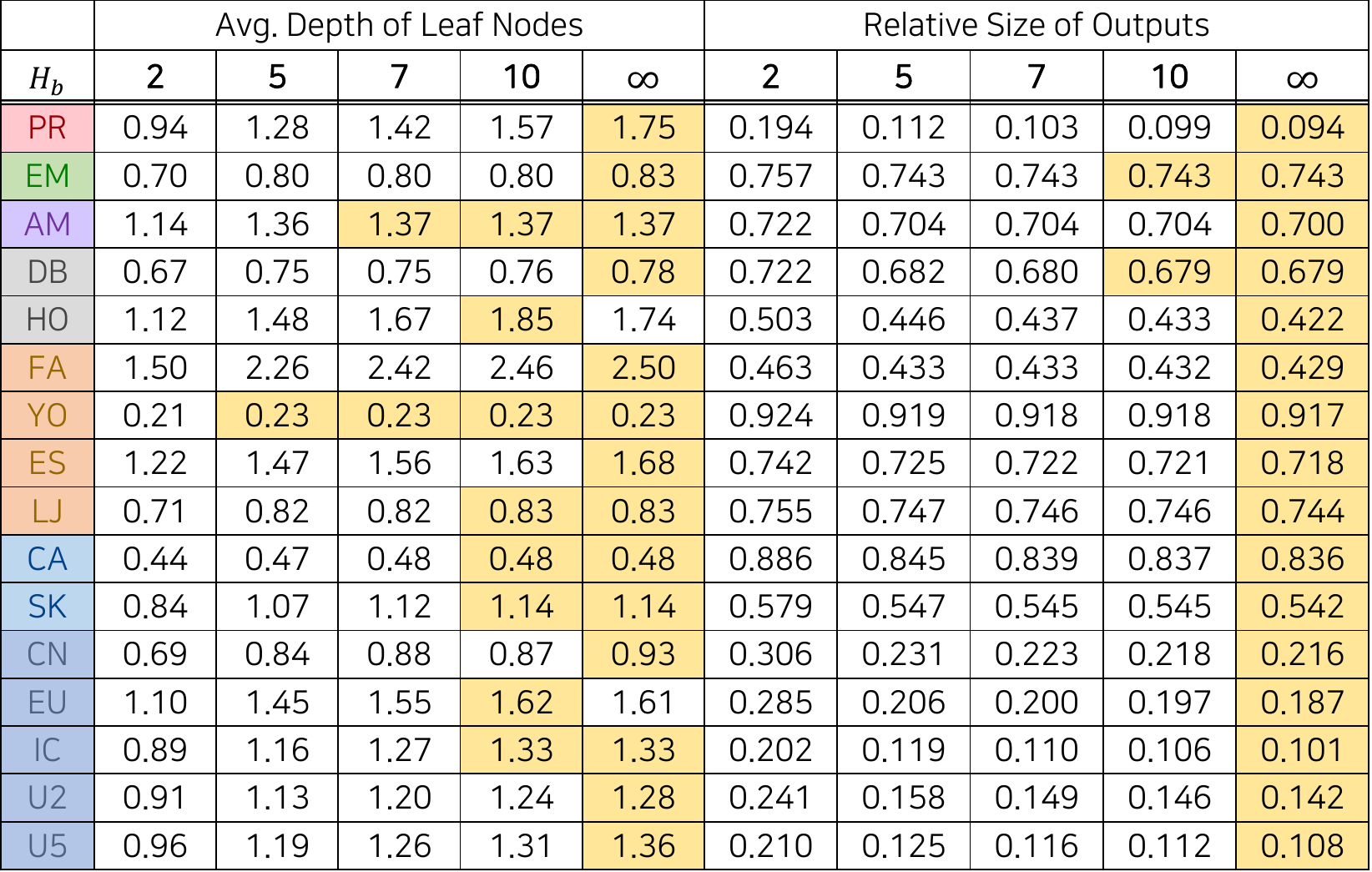} \\
\end{table}

\section{Related Work}
\label{sec:related}
Graph summarization has been studied extensively from the viewpoint of compression techniques~\cite{navlakha2008graph,boldi2004webgraph, fan2012query, lim2014slashburn, lee2020ssumm, shin2019sweg, ko2020incremental, koutra2014vog, khan2015set,beg2018scalable},
query processing~\cite{fan2012query, riondato2017graph, lefevre2010grass}, visualization~\cite{dunne2013motif, lin2008summarization, shah2015timecrunch, shen2006visual}. A survey~\cite{liu2018graph} covers these topics in detail. Below, we focus on previous works most closely related to our work: (1) lossless graph summarization, (2) lossy graph summarization, and (3) stochastic block models.

\smallsection{Lossless Graph Summarization:}
The lossless graph summarization problem (see Sect.~\ref{sec:model:prev}) was first proposed in \cite{navlakha2008graph}, and a number of algorithms have been proposed.
\randomized~\cite{navlakha2008graph} repeats (1) randomly selecting a node $u$ and (2)  merging $u$ with a node in the 2-hop neighborhood of $u$ so that the encoding cost reduces most. \sags~\cite{khan2015set} rapidly selects nodes to be merged using locality sensitive hashing instead of choosing them after comparing the reduction in the encoding cost, which is computationally expensive. %
\sweg~\cite{shin2019sweg} first groups promising node pairs to be merged using min-hashing, and within each group, it rapidly chooses pairs using Jaccard similarity instead of the actual reduction in the encoding cost.
\textsc{MoSSo}~\cite{ko2020incremental} is an online algorithm for summarizing a fully dynamic graph stream.
In response to each edge addition or deletion in the input graph, it performs an incremental update of the maintained output representation while maintaining compression rates comparable to those of offline algorithms.
Recall that these algorithms were compared with \method in terms of compression rates and speed in Sect.~\ref{sec:expr:compact} and \ref{sec:expr:speed}.

\smallsection{Lossy Graph Summarization:} 
A lossy variant of the graph summarization problem \cite{navlakha2008graph} is to find the most concise representation \summary without changing more than $\epsilon$ of the neighbors of each node in the decompressed graph. \apx~\cite{navlakha2008graph}, and \sweg~\cite{shin2019sweg} were proposed for the variant.
Other variants \cite{lefevre2010grass,lee2020ssumm,zhou2021dpgs} consider a graph representation model without corrections (i.e., $C^+$ and $C^-$) but with weight edges in $P$, and it is to find the most accurate representation of the input graph using a given number of supernodes.
For this variant, \kGs~\cite{lefevre2010grass} repeatedly merges node pairs from sampled candidate node pools that best minimize the entry-wise \lone-norm between the adjacency matrix of the input graph and that reconstructed from the output representation. For the same variant, \ssl~\cite{riondato2017graph} provides an approximation guarantee in terms of the $\ell_{p}$ reconstruction error with respect to the input graph, using geometric clustering.
\ssumm~\cite{lee2020ssumm} summarizes the original graph within a given number of bits while balancing the size of the output and its accuracy by adopting the minimum description length principle~\cite{rissanen1978modeling}.

\smallsection{Stochastic Block Model:}
Our work is related to stochastic block models (SBMs)~\cite{holland1983stochastic, karrer2011stochastic}, in the sense that they aim to group nodes with similar connectivity patterns.
Especially, there are hierarchical variants of them~\cite{clauset2008hierarchical, leskovec2008statistical, park2017fast, yan2012hierarchical}.
Since SBMs use only a single fraction to encode all edges between two groups of nodes, they tend to cause significant information loss when they are applied to real-world graphs. Thus, they are not applicable to lossless graph compression.

\section{Conclusion}
\label{sec:conclusion}
In this work, we propose the hierarchical graph summarization model, which can naturally exploit hierarchical structures, which are prevalent in real-world graphs. Then, we propose \method, a scalable algorithm for summarizing massive graphs with the new model. \method is a randomized greedy algorithm equipped with sampling, approximation, and memoization, which leads to linear scalability.
We summarize our contributions as follows:
\begin{itemize}[leftmargin=*]
	\item \textbf{New Graph Representation Model}: We propose the hierarchical graph summarization model, which succinctly and naturally represents pervasive hierarchical structures in real-world graphs.
	It generalizes the previous graph summarization model with more expressive power (Theorem~\ref{theorem:nk}).
	\item \textbf{Fast and Effective Algorithm}: We propose \method for finding parameters of our model that concisely and exactly represent the input graph. \method gives up to $\mathbf{29.6\%}$ more concise representations than its best competitors (Fig.~\ref{fig:size}).
	Moreover, \method scales linearly with the size of the input graph (Fig.~\ref{fig:scalability}) and successfully summarizes a graph with up to $\mathbf{0.8}$ billion edges. 
	\item \textbf{Extensive Experiments}: Through comparison with four state-of-the-art graph summarization methods on $16$ real-world graphs, we demonstrate the advantages of \method.
\end{itemize}
\noindent \textbf{Reproducibility}: The source code and the datasets are available at \url{https://github.com/KyuhanLee/slugger}. %

  {\small \smallsection{Acknowledgements:} This work was supported by National Research Foundation of Korea (NRF) grant funded by the Korea government (MSIT) (No. NRF-2020R1C1C1008296) and Institute of Information \& Communications Technology Planning \& Evaluation (IITP) grant funded by the Korea government (MSIT) (No. 2019-0-00075, Artificial Intelligence Graduate School Program (KAIST)).}

\section{Appendix: Proofs}
\label{sec:appendix:proofs}
\subsection{Proof of Theorem~\ref{theorem:nk}}
\label{sec:appendix:example}
To prove Theorem~\ref{theorem:nk}, we first prove the following lemma.
\begin{lemma}[]\label{theorem:nk:lemma2}
	If a supernode $A \in S$ contains more than or equal to $8k$ subnodes, $(A,B)$ $\in$ $P$ for every $B$ $\in$ $S$.
\end{lemma}

\begin{proof}
	For every subnode $u$, the number of subnodes that are not directly connected to $u$ is exactly $2k$. Thus, for every subnode $u \in B$, the number of all subedges that connect $u$ and any subnode in $A$ is at least $|A| - 2k -  1$, if $A = B$, or $|A| - 2k$, otherwise.
	
	Let $T_{AB}$ be the set of all possible edges between supernodes $A$ and $B$, and let $E_{AB}$ be the set of edges between supernodes $A$ and $B$. Then $(A,B)$ should be contained in $P$ where $|T_{AB}| - |E_{AB}| + 1 < |E_{AB}|$.
	
	For $A = B$, the statement holds because
	\begin{align*}
	|T_{AA}| - |E_{AA}| + 1 &\leq \frac{|A|(|A| - 1)}{2} - \frac{|A|(|A| - 2k - 1)}{2} + 1 \\
	&= k|A| + 1 \leq (k+1)|A| < \frac{6k-1}{2}|A| \\
	&\leq \frac{|A|(|A| - 2k - 1)}{2} \leq |E_{AA}|
	\end{align*}
	
	for $k \geq 1$. Similarly, for $A$ $\neq$ $B$, 
	\begin{align*}
	|T_{AB}| - |E_{AB}| + 1 &\leq |A||B| - (|A| - 2k)|B| + 1 \\
	&< 6k|B| \leq (|A| - 2k)|B| \leq |E_{AB}| %
	\end{align*}
	
	holds since $|B| \geq k \geq 1$.
\end{proof}

\begin{proof} (Proof of Theorem~\ref{theorem:nk})
	The size of ${V \choose 2} \setminus E$ is exactly $nk^2$. Suppose $U$ be the union of supernodes whose size is greater than or equal to $8k$. Then the subset $D = \{d \in {V \choose 2} \setminus E \mid d \cap U \neq \emptyset \}$ of ${V \choose 2} \setminus E$ is also a subset of $C^-$ by the previous lemma.
	
	Assume $G$ can be represented with $O(nk)$ edges by the previous graph summarization model. Then, $|D|$ should be $O(nk)$, %
	and the set $\left\{d \in {V \choose 2} \setminus E \mid d \subseteq V \setminus U \right\}$ should contain $\Theta(nk^2)$ distinct elements. Since every subnode has degree $(n - 2)k - 1$, $|V \setminus U|$ should be $\Theta(\frac{nk^2}{2k}) = \Theta(nk)$. Also, the number of supernodes containing at most $8k-1$ subnodes should be $\Omega(\frac{nk}{8k-1}) = \Omega(n)$.

	Hence, there are at least $\Omega(n^2) - nk^2$ unordered supernode pairs $\{A,B\}$ such that the number of edges between the supernodes $|E_{AB}|$ is nonzero and equivalent to the number of all possible edges $|T_{AB}|$. For those pairs, the minimum encoding cost between $A$ and $B$ is exactly $1$ because $|E_{AB}|$ is nonzero. Therefore, the overall encoding cost should be at least $\Omega(\frac{n(n-1)}{2} - nk^2) \in \Omega(n^{1.5})$, which contradicts the assumption.
\end{proof}

\subsection{Proof of Lemma~\ref{lemma:3hop}}
\label{sec:appendix:3hop}
\begin{proof}
	Suppose two root nodes $A\neq B\in R$ that are $3$ or more distance away from each other are merged into a new root node $A \cup B$ and $\hat{G}$ is updated from $\overline{G}$, as described Sect.~\ref{alg:merging}. Then for all $C \in S \setminus \{A, B\}$, the following equalities hold:
	\begin{align}
	& Cost^{P}_{A,B}(\overline{G})= 0,\label{eq:top1} \\
	& Cost^{P}_{A,A}(\overline{G}) +  Cost^{P}_{B,B}(\overline{G})= Cost^{P}_{A\cup B,A\cup B}(\hat{G}),\label{eq:top3} \\
	& Cost^{P}_{A,C}(\overline{G})= 0\;\; or \;\; Cost^{P}_{B,C}(\overline{G})= 0,\label{eq:top2} \\ %
	& Cost^{H}_{A \cup B}(\hat{G})= Cost^{H}_{A}(\overline{G}) + Cost^{H}_{B}(\overline{G}) + 2.\label{eq:topH}
	\end{align}
	Eq.~(\ref{eq:top2}) implies
	\begin{equation}
	Cost^{P}_{A,C}(\overline{G}) + Cost^{P}_{B,C}(\overline{G}) =  Cost^{P}_{A \cup B,C}(\hat{G}). \label{eq:top4} %
	\end{equation}
	The cost before the merger can be divided as follows:
	\begin{align}
	& Cost_{A}(\overline{G})+Cost_{B}(\overline{G})- Cost^{P}_{A, B}(\overline{G}) \;\;\;\;\;\;\;\;\;\;\;\;\;\;\;\;\; \label{3hop:final} \\
	&= Cost^H_A(\overline{G}) + Cost^H_B(\overline{G})+  Cost^{P}_{A, A}(\overline{G})+Cost^{P}_{B, B}(\overline{G}) \nonumber \\
	& + Cost^{P}_{A, B}(\overline{G})+ \sum_{C \in S \setminus \{A,B\}}(Cost^{P}_{A,C}(\overline{G}) + Cost^{P}_{B,C}(\overline{G})). \nonumber
	\end{align}
	Thus, Eqs.~(\ref{eq:top1}), (\ref{eq:top3}), (\ref{eq:topH}), (\ref{eq:top4}), and (\ref{3hop:final}) imply
	\begin{equation}
	Cost_{A \cup B}(\hat{G}) = Cost_{A}(\overline{G}) + Cost_{B}(\overline{G}) + 2, \label{eq:result}
	\end{equation}
	Hence, Eqs.~(\ref{eq:top1}), and (\ref{eq:result}) imply
	Eq.~(\ref{eq:3hop}).
\end{proof}

\label{sec:appendix:analysis}

\subsection{Proof of Lemma~\ref{lemma:time:candigen}}

\begin{proof}
Computing the hash values of all subnodes takes $O(|V|)$ time.
For each root node, computing its shingle value requires comparing the hash values of all subnodes adjacent to the subnodes contained in the root node.
Computing the shingle values of all root nodes can be performed while iterating over all edges once, 
and thus its time complexity is $O(|E|)$.
Lastly, grouping root nodes according to their shingle values takes $O(|V|)$ time.
Hence, the overall time complexity of the candidate generation step is $O(|V|+|E|)=O(|E|)$.
\end{proof}

\subsection{Proof of Lemma~\ref{lemma:time:merge}}

\begin{proof}
To compute the saving (i.e., Eq.~\eqref{eq:relative_saving}) between two distinct supernodes $X$ and $Y$, all the edges incident to one of the descendants of $X$ or $Y$ need to be accessed, and the number of such edges is bounded by $|\{(u,v) \in E \mid u \in X \cup Y, v \in V \}|=O(\sum_{v \in X} deg(v) + \sum_{v \in Y} deg(v))$.
For each candidate root node set $D \in C_t$, in an iteration, the savings between a supernode $A \in D$ and the other nodes in $D$ need to be computed, and it takes $O(|D| \cdot \sum_{v \in A} deg(v) + \sum_{Y \in D - A} \sum_{v \in Y} deg(v))=O\left(|D| \cdot \sum_{X \in D} \sum_{v \in X} deg(v)\right)$ time. 
Thus, processing $D$, which requires at most $|D|-1$ iterations, takes $O(|D|^2 \cdot \sum_{X \in D} \sum_{v \in X} deg(v))$ time. Since $|D|$ is at most a constant, as described in Sect.~\ref{alg:candiate}, the overall time complexity of the merging step is
$$
O\left(\sum_{D \in C_t} \sum_{v \in D} deg(v)\right) = O\left(\sum_{v \in V} deg(v)\right) = O(|E|). \;\;\; \qedhere
$$
\end{proof}

\subsection{Proof of Lemma~\ref{lemma:time:pruning}}
\label{sec:appendix:time:pruning}
\begin{proof}
The worst-case time complexity of Step $1$ and Step $2$ is $O(|H| + |V|) = O(|E|)$. 
Let $P_{AB}$ be the set of $p$-edges and $n$-edges %
between $S_A$ and $S_B$, $E_{AB}$ be the set of subedges between $S_A$ and $S_B$, and $T_{AB}$ be the set of all possible subedges between $S_A$ and $S_B$. 
Then for every unordered root node pair $\{A,B\}$ in the last pruning step, %
computing the encoding cost using the previous model requires $O(|E_{AB}|)$ and the corresponding cost using the current representation requires $O(|P_{AB}|)$ since we can compute the cost by retrieving the $p$-edges and $n$-edges.
The worst-case time complexity of updating the encoding is $O(|P_{AB}| + \min\{|T_{AB}| - |E_{AB}| + 1, |E_{AB}|\})$, since it requires removing all $p$-edges and $n$-edges between $A$ and $B$ and adding edges based on the new encoding. Hence, the overall time complexity of the pruning step is $O(|P^+|+|P^-|+|E|) = O(|E|)$.
\end{proof}

\subsection{Proof of Lemma~\ref{lemma:space}}
\label{sec:appendix:space}
\begin{proof}
During the whole process, \method requires the input graph and the membership of leaf nodes to root nodes, so an additional $O(|V|+|E|)$ space is needed.
For the candidate generation step, $O(|V|)$ space is required to store both the hash and shingle values of every node. Also, storing generated candidate sets requires $O(|V|)$ space.
For memoization, \method requires constant space since the number of cases is constant.
For the merging step, \method requires connections between every root node pair whose descendants are connected in $G$, in addition to all three types of edges. Since the number of such connections and the number of edges are $O(|E|)$, the space complexity for the merging step is $O(|E|)$.
For the pruning step, \method only requires all three types of edges, and thus the space complexity is $O(|E|)$.
Therefore, the overall space complexity is $O(|V|+|E|)=O(|E|)$.
\end{proof}

\label{sec:appendix:stderr}

\section{Appendix: Algorithmic Details}

\subsection{Pseudocode for Pruning}
\label{sec:appendix:prune}

Algorithm~\ref{alg:prune} gives the first two pruning steps in \method. 

\begin{algorithm}[h!]
	\SetAlgoLined
	\LinesNumbered
	\KwData{(a) input graph: \original \\
		\quad (b) curr. hier. graph summ. model: \sG  \\
		\quad (c) set of root nodes: $R$ \\
	}
	\KwResult{updated \sG}
	\caption{Pruning Step\label{alg:prune}}
	$Q$ $\leftarrow$ $R$  \Comment*[f]{$H(X) = \{Y : (X,Y) \in H \wedge Y \subseteq X \}$} \\ %
	\While{$Q \neq \emptyset$}{\label{prune:line:step1} 
	\vspace{-4mm}
	\Comment*[f]{Step.~$1$} \\
	    pick and remove a random node $A$ from $Q$  \\
		\If{$\nexists B$ s.t. $(A, B) \in P^+ \cup P^-$ and $H(A) \neq \emptyset$}{ 
			$S$ $\leftarrow$ $S \setminus \{A\}$ \\
			\eIf{$A \in R$}{
			    $R$ $\leftarrow$ $(R \setminus \{A\}) \cup H(A)$
			}{
			    $Pr$ $\leftarrow$ the parent of the node $A$ \\
			    $H$ $\leftarrow$ $H \cup \{(Pr, X) : X \in H(A)\}$
			}
			$H$ $\leftarrow$ $H \setminus \{(A,X) : X \in H(A)\}$ \\
		}
		$Q$ $\leftarrow$ $Q \cup H(A)$  \\
	}
	$Q$ $\leftarrow$ $R$  \\
	\While{$Q \neq \emptyset$}{\label{prune:line:step2} 
	\vspace{-3.7mm}
	\Comment*[f]{Step.~$2$} \\
		pick and remove a random root node $A$ from $Q$  \\
		\If{$\exists! B$ s.t. $(A, B) \in P^+ \cup P^-$ and $A \neq B$}{
			$D$ $\leftarrow$ $\{(C, B) : C \in H(A)\}$  \\
			\eIf{$(A,B) \in P^+$}{
				$P^+$ $\leftarrow$ $(P^+ \setminus \{(A,B)\}) \cup (D \setminus P^-)$ \\
				$P^-$ $\leftarrow$ $P^- \setminus D$  \\
			}{
				$P^-$ $\leftarrow$ $(P^- \setminus \{(A,B)\}) \cup (D \setminus P^+)$ \\
				$P^+$ $\leftarrow$ $P^+ \setminus D$ \\
			}
			$S$ $\leftarrow$ $S \setminus \{A\}$ \\
			$H$ $\leftarrow$ $H \setminus \{(A,X) : X \in H(A)\}$ \\
			$R$ $\leftarrow$ $(R \setminus \{A\}) \cup H(A)$ \\
			$Q$ $\leftarrow$ $Q \cup H(A)$ \\
		} 
	}
	\textbf{return} \sG
\end{algorithm}

\subsection{Pseudocode for Partial Decompression}
\label{sec:appendix:query}

Algorithm~\ref{alg:query} shows  how to retrieve the neighbors of a given node from our model without decompressing the entire model. 
We measured the average time taken for retrieving the neighbors of a node in the hierarchical summary graphs obtained by \method from the $16$ considered datasets. 
The time was less than $15$ microseconds on all hierarchical summary graphs, and it was less than $4$ microseconds on $12$ hierarchical summary graphs.
The time was strongly correlated with the average depth of leaf nodes in hierarchy trees. The Pearson correlation coefficient between them was about $0.82$.
Full results including the time taken on summary graphs obtained by \method and \sweg can be found in \cite{supple}.

\subsection{Graph Algorithms on a Hierarchical Summary Graph}
\label{sec:appendix:algorithms}

A variety of unweighted graph algorithms, including DFS, PageRank, and Dijkstra's, access the input graph only by retrieving the neighbors of a node.
For example, line~\ref{alg:dfs:access} of Algorithm~\ref{alg:dfs} and line~\ref{alg:pagerank:access} of Algorithm~\ref{alg:pagerank}, where the neighbors of a node are retrieved, are the only step where the input graph is accessed in DFS and PageRank.
Even if the input graph is represented as a hierarchical summary graph, the neighbors of a given node can be retrieved efficiently through partial decompression, as discussed in Sect.~\ref{sec:appendix:query}.
Thus, such graph algorithms, with no or minimal changes, can run directly on a hierarchical summary graph by partially decompressing it on-the-fly. We report the running time of four algorithms (BFS, PageRank, Dijkstra's, and triangle counting) executed on summary graphs obtained by \method and \sweg in \cite{supple}.

\begin{algorithm}[h]
	\SetAlgoLined
	\LinesNumbered
	\KwData{
	    (a) hier. graph summ. model: \sG\\
		\hspace{9.7mm}  (b) a subnode $v \in V$ \\
	}
	\KwResult{the set $N_v$ of $v$'s one-hop neighbors in $G$}
	\caption{Partial Decompression of Hierarchical Graph Summarization Models \label{alg:query}}
	node $\leftarrow$ $v$ \\ %
	$M$ $\leftarrow$ an empty stack \\ %
	$count$ $\leftarrow$ an empty dictionary \\
	\While{node $!= null$}{
		M.push(node) \\
		node $\leftarrow$ parent of node \\
	}
	\While{$!M.isEmpty()$}{
		$X$ $\leftarrow$ M.pop() \\
		\ForEach{$A$ $\in$ $\{Y:(X, Y) \in P^+\}$}{
		    \ForEach{$u$ $\in$ $A$}{
		        $count(u)$ $\leftarrow$ $count(u) + 1$ \\
		    }
		}
		\ForEach{$A$ $\in$ $\{Y:(X, Y) \in P^-\}$}{
			\ForEach{$u$ $\in$ $A$}{
		        $count(u)$ $\leftarrow$ $count(u) - 1$ \\
		    }
		}
	}
	$N_v$ $\leftarrow$ $\emptyset$ \\
	\ForEach{$u$ $\in$ $M - v$}{
	    \If{$count(u) = 1$}{
	        $N_v$ $\leftarrow$ $N_v \cup \{u\}$
	    }
	}
	\algorithmicreturn{~$N_v$}
\end{algorithm}

\begin{algorithm}[h!]
	\SetAlgoLined
	\LinesNumbered
	\KwData{
	(a) graph: \original, (b) current node: $v$, \\
	\hspace{9.7mm} (c) set of visited node: $M$ \\
	}
	\KwResult{updated set of visited node $M$}
	\caption{Depth First Search (DFS)\label{alg:dfs}}
	\For{$\mathbf{each}$ node $u\in \blue{getNeighbors(v)}$\label{alg:dfs:access}}{
	    \If{$u \notin M$}{
	        $M$ $\leftarrow$ $M \cup \{u\}$; \ DFS($u$, $M$)
	    }
	}
	\textbf{return} $M$
\end{algorithm}

\begin{algorithm}[h!]
	\SetAlgoLined
	\LinesNumbered
	\KwData{
	(a) undirected graph: \original \\
	\hspace{9.7mm}  (b) damping factor: $d$, (c) iteration number: $T$ \\
	}
	\KwResult{PageRank vector $r^{new} \in 	\mathbb{R}^{|V|}$}
	\caption{PageRank~\cite{page1999pagerank}\label{alg:pagerank}}
	$r^{new} \leftarrow \frac{1}{|V|}\cdot \mathbf{1}$ \\
		\vspace{-4mm}
	\Comment*[f]{$\mathbf{1}=$ one vector of size $|V|$} \\
	\While{T > 0}{
    	$r^{old} \leftarrow r^{new}$; \ $r^{new} \leftarrow \mathbf{0}$ \\
	    \For{$\mathbf{each}$ node $u\in V$}{
                \For{$\mathbf{each}$ node $w \in \blue{getNeighbors(u)}$ \label{alg:pagerank:access}}{
                    $r^{new}_{w}$ $\leftarrow$ $r^{new}_{w} + \frac{1}{|N_{u}|} r^{old}_{u}$
                }
	    }
        $r^{new}$ $\leftarrow$ $d \cdot r^{new} + (1 - d \cdot \sum_{v\in V} r^{new}_{v})\cdot \frac{1}{|V|} \cdot \mathbf{1}$\\
        $T \leftarrow T-1$ \\
	}
	\textbf{return} $r^{new}$
\end{algorithm}

\newpage
\bibliographystyle{IEEEtran}
\bibliography{reference}

\end{document}